\documentclass[acmsmall,screen,nonacm]{acmart}

\setcopyright{none}
\renewcommand\footnotetextcopyrightpermission[1]{}

\usepackage{graphicx}
\usepackage{colortbl}
\usepackage{microtype}
\usepackage{balance}
\usepackage{mathtools}
\usepackage{multirow}
\usepackage{array}

\begin{document}

\title{Enhancing the Power of Polyhedral-Based Optimizations with Coordinate-Based Hill Climbing}

\author{Gaurav Verma}
\orcid{0000-0002-8820-6016}
\affiliation{%
  \institution{Stony Brook University}
  \city{Stony Brook}
  \country{USA}
}
\email{gaurav.verma@stonybrook.edu}

\author{Michael Canesche}
\orcid{0000-0001-7882-0787}
\affiliation{%
  \institution{Cadence}
  \city{Belo Horizonte}
  \country{Brazil}
}
\email{canesche@cadence.com}

\author{Fernando Magno Quint\~{a}o Pereira}
\orcid{0000-0002-0375-1657}
\affiliation{%
  \institution{UFMG}
  \city{Belo Horizonte}
  \country{Brazil}
}
\email{fernando@dcc.ufmg.br}

\begin{abstract}
This paper describes our experience extending the polyhedral compiler Pluto with 
a lightweight, coordinate-wise hill-climbing tuner that adjusts numeric 
transformation parameters—such as tile sizes and thread-block dimensions—after 
Pluto selects the kernel's loop structure. To ensure fast convergence and escape 
local minima, hill climbing is augmented with two techniques: expanded 
neighborhood exploration and a shortest-hop refinement phase. On x86 and ARM 
CPUs, tuned kernels outperform Pluto's default configuration (1.06–1.28× 
geometric mean speedup across 11 benchmarks) and static optimizers (Clang -O3, 
Polly, IOOpt), reaching performance competitive with the AutoTVM autotuner at 
substantially lower search cost. Applying the same technique to GPU thread-block 
allocation on an NVIDIA A100 yields 5.5–8.5\% improvement over default 
configurations. These results position post-optimization parameter tuning as a 
practical middle ground between fixed-cost-model polyhedral compilation and full 
autotuning.
\end{abstract}



\maketitle

\section{Introduction}
\label{sec_intro}

The polyhedral model is a mathematical framework that optimizes loops in programs by representing their iteration space as a polyhedron (a geometric shape in multi-dimensional space)\footnote{Notions of iteration and data space appeared independently in several works during the 80s and 90s~\cite{Wol87, Wolf91, Feautrier92b, Feautrier92a, Amarasinghe93, Irigoin88, Quinton89, Fortes84, Lam89, Rajopadhye86, Rajopadhye86b, Schreiber90, Pugh92, Ramanujam90,Wolf91}, eventually leading to the theory known today as the \textit{Polyhedral Model}.}.
The theory of polyhedra is important because it allows the systematic representation and optimization of loop-intensive code.
By modeling loop iterations as points in a high-dimensional geometric space, polyhedra enable precise analyses of data dependencies, allowing compilers to safely apply complex transformations such as loop interchange, tiling, and fusion to improve cache utilization and parallelism~\cite{Abdelaal21, Zinenko17, Gupta99, Tollenaere22, Zhang11, Oliveira23, Bondhugula08, Pouchet09}.

Given the tremendous popularity of deep-learning models, plus the fact that these models are primarily implemented as nests of loops, one might expect the polyhedral model to be enjoying its golden age.
However, modern kernel optimizers like Ansor~\cite{Zheng20}, Atlas~\cite{Whaley01}, AutoTVM~\cite{Chen18}, cuDNN~\cite{Chetlur14}, Halide~\cite{Kelley13}, OpenTuner~\cite{Ansel14}, Spiral~\cite{Franchetti18}, TensorComprehensions~\cite{Vasilache18}, and Tira\-misu~\cite{Baghdadi20} often take a different approach.
Rather than relying on purely static analyses, they perform an {\it empirical exploration} of the optimization space.
These tools employ techniques such as evolutionary search, gradient-based methods, and machine learning to sample the universe of high-performance kernel implementations.

This empirical approach enables compilers to fine-tune performance for specific hardware platforms by capturing hardware characteristics---such as cache hierarchies, vectorization capabilities, and memory bandwidth---that static models like the polyhedral approach may not fully account for.
By collecting performance data of multiple versions of the kernel, and iteratively refining them, these compilers can adapt to the intricacies of the target architecture.
This dynamic search process is often referred to as {\it auto-tuning}.
While auto-tuning is more hardware-aware than polyhedral-based optimizations, it requires the costly execution of multiple versions of the same program.

\paragraph{The Contribution of this Work}

This paper presents a practical methodology that enhances static polyhedral compilation with lightweight, post-generation parameter tuning. Unlike full autotuners that explore both kernel shape and numeric parameters empirically, we use polyhedral analysis to determine a promising kernel ``shape'' (the ordering of loops and memory access patterns) and then apply coordinate-wise hill climbing exclusively to numeric parameters such as tile sizes, unrolling factors, and thread-block dimensions. This hybrid approach offers three specific contributions:

\begin{description}

\item[A Focused Tuning Problem.] Prior work on combining polyhedral analysis with search-based optimization has explored large, combinatorial spaces that include scheduling decisions, fusion/fission choices, and transformation sequences~\cite{Papenhausen18,Trifunovic11,Pouchet08}. While powerful, these approaches incur significant search costs due to the size of the space they explore. In contrast, we observe that once a polyhedral compiler has committed to a kernel shape, the remaining optimization space (numeric parameters of already-selected transformations) is substantially smaller and amenable to lightweight local search. Our contribution is not a more expressive search, but rather a demonstration that this focused, post-hoc tuning recovers most of the available performance at a fraction of the cost of broader exploration.

\item[Static Bootstrapping of the Search Space.] Hill climbing and other gradient-based search methods require a well-defined starting point and search space to be effective~\cite{Canesche24,Li24}. Prior applications of hill climbing to kernel optimization have relied on human-provided initial configurations or on seeding from other autotuning algorithms~\cite{Canesche24,Canesche24b}. We show that polyhedral analysis alone can provide both the search space (by fixing the kernel shape) and a competitive starting point (Pluto's default parameters), eliminating the need for manual intervention or expensive pre-tuning. This addresses a question left open by \citet{Canesche24} and \citet{Zhao24}: how to identify a promising optimization space for gradient-based tuning without empirical sampling.

\item[Practical Effectiveness Across Architectures and Optimizations.] We demonstrate that our lightweight tuning approach generalizes beyond CPU tiling to GPU thread-block allocation, a qualitatively different optimization with different performance trade-offs (register pressure vs. occupancy). Across 11 CPU benchmarks and 10 GPU benchmarks, our technique consistently improves upon static compilers and achieves performance competitive with AutoTVM~\cite{Chen18}, a full autotuner, while requiring substantially less search time. These results establish post-optimization parameter tuning as a cost-effective parameter tuner.

\end{description}

\paragraph{Summary of Results}

As Section~\ref{sec_eval} shows, the proposed parameter tuning approach has been applied to two parametric compiler optimizations, each targeting a different execution environment:

\begin{description}

\item[Loop Tiling:] a combination of strip-mining and loop interchange that reorganizes iteration spaces into blocks called tiles. This optimization, implemented in Pluto~\cite{Bondhugula08}\footnote{The ideas discussed in this paper should be applicable to other polyhedral tools. We chose Pluto because a recent comparison of polyhedral compilers, including Polly, PoCC, and PPCG, shows that it achieves some of the highest observed speedups~\cite{Thangamani24}.}, uses tile sizes as optimization parameters.

\item[Thread Blocking:] a GPU optimization that partitions threads into blocks, trading register pressure for hardware occupancy. The number of threads per block is typically fixed by the developer; in our approach, it is exposed as a tunable parameter.

\end{description}

Our evaluation on x86 and ARM CPUs shows that hill climbing, augmented with expanded neighborhoods and a shortest-hop refinement phase, consistently improves upon Pluto's default configuration, achieving geometric mean speedups of 1.06--1.28x across 11 benchmarks. Tuned kernels outperform static optimizers such as \texttt{clang -O3}, Polly, and IOOpt~\cite{Olivry21}, and match the performance of AutoTVM~\cite{Chen18} at substantially lower search cost. Although search overhead remains significant (40--75x relative to compilation time), it compares favorably to full autotuners that require orders of magnitude more samples.
On GPU, the same technique improves thread-block allocation on an NVIDIA A100 by 5.5--8.5\% over default configurations, with search times that are a fraction of exhaustive exploration. These results demonstrate that post-optimization parameter tuning offers a practical middle ground between fixed-cost-model polyhedral compilation and full autotuning.

\section{Optimal Configurations of Polyhedral Optimizations}
\label{sec_ovf}

The polyhedral model treats each loop iteration as a point in the loop's {\it iteration space}.
The model uses inequalities to define the bounds and dependencies between these iterations.
This combination of bounds and dependencies provides a framework that enables loop transformations such as tiling, fusion, and interchange.
Tools that utilize the polyhedral model to optimize programs, such as Pluto~\cite{Bondhugula08} or Polly~\cite{Grosser12}, generally work in three phases:
\begin{description}
\item [Static analysis:] The polyhedral optimizer analyzes the iteration space to determine how loops interact with each other and with data.

\item [Rewriting:] The optimizer applies transformations like tiling or loop interchange to improve data locality and parallelism.

\item [Code generation:] The transformed code is generated.
\end{description}
Example~\ref{ex_matrix} illustrates these notions of iteration space and dependencies.

\begin{figure}[ht]
\includegraphics[width=\columnwidth]{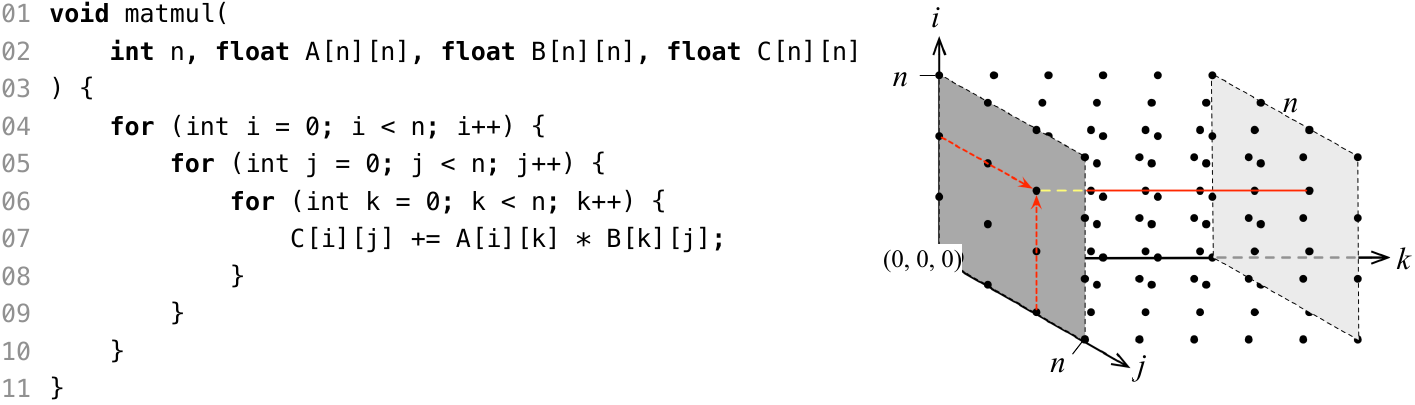}
\caption{The implementation of basic matrix multiplication,
used to illustrate the effects of polyhedral transformations, and the iteration space of this kernel, showing loop dependencies along the K dimension.}
\Description{The implementation of basic matrix multiplication,
used to illustrate the effects of polyhedral transformations.}
\label{fig_matrix_multiplication}
\end{figure}

\begin{example}
\label{ex_matrix}
Figure~\ref{fig_matrix_multiplication} shows a basic implementation of matrix multiplication, $C = A \times B$, written in the C programming language.
Each loop forms a dimension of the iteration space.
Figure~\ref{fig_matrix_multiplication} represents this set of all valid loop iterations as a 3D space with \texttt{i}, \texttt{j}, and \texttt{k} forming the axes.
Each point in this 3D space corresponds to a specific combination of loop indices $(\mathtt{i}, \mathtt{j}, \mathtt{k})$ for which an operation is performed.
The iteration space is bounded by the loop limits.
For instance, for \texttt{i}, the bounds are $0 \leq \mathtt{i} < \mathtt{n}$.
The polyhedral model captures these bounds as inequalities that form a polyhedron in a higher-dimensional space.
The model also captures dependencies between iterations.
In Figure~\ref{fig_matrix_multiplication}, there is no dependency between different iterations of \texttt{i} and \texttt{j}, but there is a dependency within the \texttt{k} loop because
\texttt{C[i][j]} is updated in every iteration of \texttt{k}.
\end{example}

\subsection{Polyhedral Transformations}
\label{sub_poly_transf}

The polyhedral model enables optimizations like loop interchange and tiling.
This theory guarantees that such transformations preserve the program's correctness by ensuring that memory dependencies across different loop iterations are respected.
Once the dependencies are captured, the model can safely apply transformations like strip mining (breaking loops into smaller blocks) and loop interchange (reordering loops to improve data locality).
Example~\ref{ex_opt} illustrates how these two transformations enable {\it loop tiling}.

\begin{figure}[ht]
\includegraphics[width=\columnwidth]{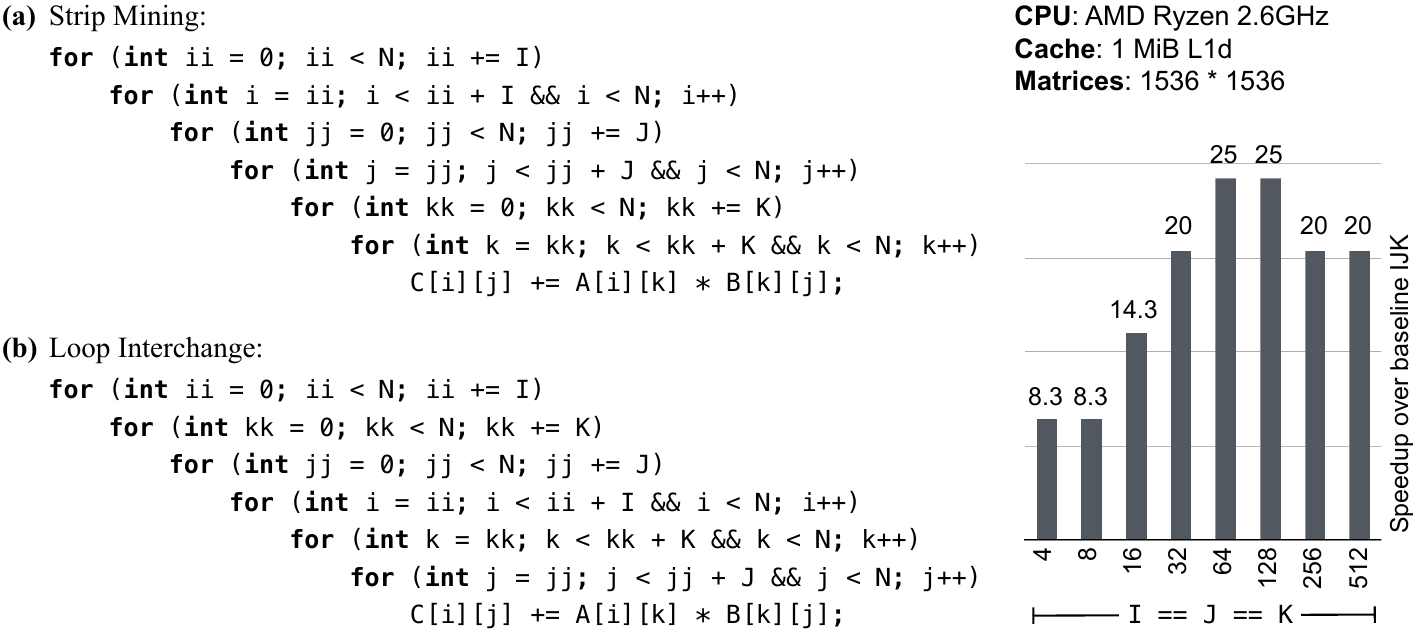}
\caption{A tiled version of the matrix multiplication algorithm (see Figure~\ref{fig_matrix_multiplication}).
This version results from the application of two compiler transformations: strip mining and loop interchange.
The bar plot shows the speedup of the tiled version over the implementation in Figure~\ref{fig_matrix_multiplication} on a commodity AMD Ryzen.}
\Description{Matrix multiplication after loop interchange and tiling.}
\label{fig_tiled_mm_k}
\end{figure}

\begin{example}
\label{ex_opt}
Figure~\ref{fig_tiled_mm_k} shows the code that results from manually applying strip mining and loop interchange to the program in Figure~\ref{fig_matrix_multiplication}.
Strip mining partitions the loops into smaller blocks.
Once combined with interchange, these blocks form {\it tiles}.
Looping over such tiles, instead of over the entire iteration space, enhances cache locality.
The optimized program runs faster: on an AMD Ryzen at 2.6GHz, the program in Figure~\ref{fig_matrix_multiplication} (b), compiled with gcc -O3, with $1024 \times 1024$ matrices, might take about 0.1 seconds to finish, whereas the na\"{i}ve version in Figure~\ref{fig_matrix_multiplication} takes more than 2 seconds.
\end{example}

\subsection{Parameter Tuning}
\label{sub_param_tuning}

The optimizations that produced the program in Figure~\ref{fig_tiled_mm_k} rely on three parameters: \texttt{I}, \texttt{J}, and \texttt{K}, the dimensions of the three tiling windows.
The values in Figure~\ref{fig_tiled_mm_k}, $\mathtt{I} = \mathtt{J} = \mathtt{K} = 32$, are the default dimensions used by tools like Pluto.
However, they are not necessarily the best values; hence, these tools usually offer users knobs to configure them.
Example~\ref{ex_surface} shows what such fine-tuning can achieve.

\begin{figure}[ht]
\includegraphics[width=1\columnwidth]{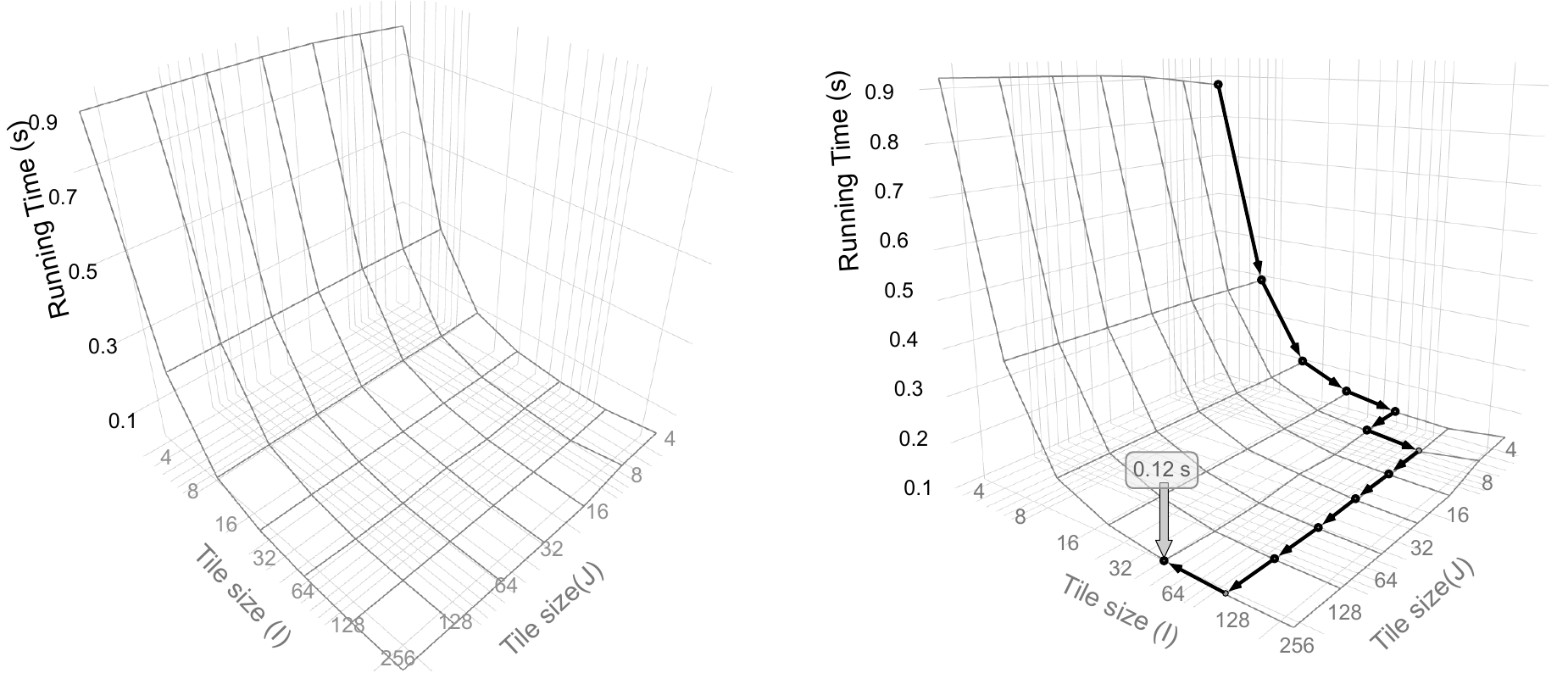}
\caption{Two different views of the search space of two tiling parameters used in Figure~\ref{fig_tiled_mm_k} on an AMD Ryzen 2.6GHz. The rightmost view highlights the path produced by hill climbing when applied to this optimization space.}
\Description{Example of search space.}
\label{fig_surface}
\end{figure}

\begin{example}
\label{ex_surface}
Figure~\ref{fig_surface} shows the search space formed by the tiling parameters \texttt{I} and \texttt{J} defined in Figure~\ref{fig_tiled_mm_k}, with $\mathtt{K} = 32$, considering square matrices of $1024 \times 1024$ elements.
The origin of this space is $(\mathtt{J} = 4, \mathtt{I} = 4)$.
The surface was produced by running the kernel in Figure~\ref{fig_tiled_mm_k} on an AMD Ryzen 2.6GHz, considering every possible combination of $(\mathtt{I} \times \mathtt{J})$, where $\{\mathtt{I}, \mathtt{J}\} \subset \{4, 8, 16, 32, 64, 128, 256\}$.
An exhaustive search on this grid reveals that the optimal kernel implementation uses
$(\mathtt{J} = 256, \mathtt{I} = 64)$, achieving a running time of 0.12 seconds.
The naive, non-optimized code seen in Figure~\ref{fig_matrix_multiplication} runs in 2.19 seconds.
Thus, optimizations, in this example, achieve a speedup of 18.25x.
\end{example}

Figure~\ref{fig_surface} shows that, in this particular example, the region extending from the origin of the search space to its optimal configuration is {\it convex}.
A convex subregion within the {\it hypersurface}\footnote{A hypersurface is a manifold of $n-1$ dimensions embedded in an $n$-dimensional space, defined by a function. In Example~\ref{ex_surface}, this function maps $(\mathtt{J}, \mathtt{I})$ to running time.} defined by a function is a subset of the hypersurface where, for any two points within the region, the straight line segment connecting them lies entirely within it.
In a convex region, any local minimum is a global minimum.
This makes optimization much simpler and more efficient, as there are no local minima traps to mislead search algorithms.
The expectation that the region containing the origin of a kernel's search space and its global optimum be convex is known as the {\it Droplet Hypothesis}~\cite{Canesche24}.
We notice that this is an ``expectation''; not a guarantee: many benchmarks that this paper analyzes yield non-convex optimization surfaces (see Figure~\ref{fig_blockDimensions} on Page~\pageref{fig_blockDimensions}).
Nevertheless, as Section~\ref{sec_eval} demonstrates, convexity is not a necessary requirement for the adjustment policy proposed in this paper to work.
However, whenever this property is present, the adjustment is guaranteed to converge to an optimal kernel configuration.

\section{Coordinate-Wise Hill Climbing over Polyhedra}
\label{sec_sol}

Our search procedure is a \textit{hill-climbing} algorithm: it repeatedly moves to a better-performing point in a discrete search space by comparing the current candidate against a set of nearby candidates, called its \textit{neighborhood}.
Each neighbor differs from the current candidate in exactly one optimization parameter (one ``coordinate''), so we call the approach \textit{coordinate-wise hill climbing}\footnote{Notice that this approach is a different search strategy from \textit{coordinate descent}, which instead fixes all but one coordinate and performs a full search (e.g., a line search) along that single coordinate before moving on to the next.}.
The essence of our search algorithm consists of four steps:

\begin{description}
\item[Initialization:] Start with an initial guess for the solution, and let it be the current kernel candidate (see Section~\ref{sub_boot}).

\item[Neighborhood Generation:]
For each iteration, generate the neighborhood of the current candidate using the definition of neighborhood from Section~\ref{sub_neigh}.

\item[Candidate Selection:] Evaluate candidates in the neighborhood and update the current candidate to the one that improves the objective function, using one of the selection strategies described in Section~\ref{sub_next_kernel}.

\item[Convergence:] The process continues until some convergence criterion (like a small enough change in the objective function or solution) is met (see Section~\ref{sub_stop}).
\end{description}

\subsection{Bootstrapping: Finding the Seed Kernel}
\label{sub_boot}

In this work, we use the Polyhedral Model to {\it bootstrap} parameter search via coordinate-wise hill climbing.
Bootstrapping, in this context, means finding a {\it Kernel Optimization Space}.
To define this concept, we note that in the Polyhedral Model, typical kernels can be written as a set of polyhedra, each defined by the loop bounds.
Example~\ref{ex_polyhedra} illustrates this approach.

\begin{example}
\label{ex_polyhedra}
In the polyhedral model, the iteration space of the loops in the matrix multiplication kernel of Figure~\ref{fig_matrix_multiplication} are described as a set of polyhedra defined by the loop bounds.
The indices \( i \), \( j \), and \( k \) form a three-dimensional space, and the bounds \( 0 \leq i < N \), \( 0 \leq j < N \), and \( 0 \leq k < N \) define a polyhedron (a cube) in this space defined as:

\[
\mathcal{P} = \{ (i, j, k) \mid 0 \leq i < N, \ 0 \leq j < N, \ 0 \leq k < N \}
\]

This 3D cube represents the entire set of iterations for the three loops. Each point in this cube corresponds to a specific combination of loop indices \( (i, j, k) \), and the computation at that point updates the corresponding element \( C[i][j] \).
\end{example}

Given the polyhedral representation of a loop, code transformations can be described as transformations of this geometric space.
In other words, because the iteration space is represented as a set of integer points, transformations can be applied by multiplying these points by a transformation matrix.
Example~\ref{ex_transform} illustrates this point.

\begin{example}
\label{ex_transform}
To perform a \textbf{loop interchange} on the representation seen in Example~\ref{ex_polyhedra}, we might swap the loops over \(j\) and \(k\).
In this case, the iteration space becomes:

\[
\mathcal{P'} = \{ (i, k, j) \mid 0 \leq i < N, \ 0 \leq k < N, \ 0 \leq j < N \}
\]

We represent this transformation with a \textbf{permutation matrix} that swaps the second and third components of the vector.
The matrix that swaps \(j\) and \(k\) is:

\[
T = \begin{pmatrix} 
1 & 0 & 0 \\
0 & 0 & 1 \\
0 & 1 & 0 
\end{pmatrix}
\]

This matrix has the following effect:

\begin{itemize}
\item The first row leaves the first coordinate \(i\) unchanged.
\item The second row places \(k\) in the second position.
\item The third row places \(j\) in the third position.
\end{itemize}

To apply the transformation to a point \(\mathbf{p}\), we multiply the point by a matrix \(\Theta\):

\[
\Theta \cdot \mathbf{p} = 
\begin{pmatrix} 
1 & 0 & 0 \\
0 & 0 & 1 \\
0 & 1 & 0 
\end{pmatrix}
\cdot
\begin{pmatrix} 
i \\ 
j \\ 
k 
\end{pmatrix}
=
\begin{pmatrix} 
i \\ 
k \\ 
j 
\end{pmatrix}
=
\mathbf{p'}
\]

This new point $\mathbf{p'}$ represents the iteration space after the loop interchange.
\end{example}

The notions that Examples~\ref{ex_polyhedra} and~\ref{ex_transform} illustrate give us the material to formalize the {\it Kernel Optimization Space}:

\begin{definition}[Kernel Optimization Space]
\label{def_space}
Let $\mathcal{P}$ be the polyhedra-based description of a computational kernel.
Let $S = \langle \Theta_1, \Theta_2, \ldots, \Theta_n \rangle$ be an ordered sequence of $n$ transformations applied to $\mathcal{P}$; hence, leading to a new polyhedra-based description $\mathcal{P}'$.
We call the sequence $S$ a kernel optimization space relative to kernel $\mathcal{P}$.
\end{definition}

\begin{example}
\label{ex_space}
Function \texttt{tiled} (Fig.~\ref{fig_tiled_mm_k}) shows one concrete kernel that inhabits the space that ensues from the following sequence of optimizations applied onto Figure~\ref{fig_matrix_multiplication}:
$S = \langle \mathtt{interchange}_{i,j}, \allowbreak
\mathtt{tile}_j, \allowbreak \mathtt{tile}_k \rangle$.
Figure~\ref{fig_surface} shows the full space, which is formed by every possible kernel that abide by the same
sequence of optimizations.
\end{example}

The beauty of polyhedral-based transformations is that they provide a composable technique to rewrite loops, plus the mechanisms to validate these transformations and to estimate their profit.
In this regard, polytope operations give us a measure of {\it memory reuse}: by comparing the dimensions of the data space (the bounds of tensors accessed by kernels) with the dimensions of the iteration space we can estimate data locality statically~\cite{Benabderrahmane10}.
These observations lead to Proposition~\ref{prop_poly}, which we state below:
\begin{proposition}
\label{prop_poly}
Out-of-the-box polyhedral analyses can be effectively used to find a good kernel optimization space.
\end{proposition}
Notice that Proposition~\ref{prop_poly} talks about the {\it kernel space} but says nothing within points within this space: this optimal point depends on hardware details.
It is not simple to augment polyhedral analyses with cost models that correctly model the many architectures onto which they can be used, as already observed by previous work~\cite{Tollenaere23}.
To achieve this goal, we resort to the ideas discussed in Section~\ref{sub_neigh}.

\subsection{The Neighborhood Function}
\label{sub_neigh}

The Search Space from Definition~\ref{def_space} has a {\it basis}: a set of parameterizable optimizations that spans the entire space.
These parameters, in the context of this paper, are natural numbers.
In other words, each point of the optimization space is a choice of values for the parameters of the optimizations that define it.
Example~\ref{ex_basis} clarifies this notion.

\begin{figure}[ht]
\includegraphics[width=\columnwidth]{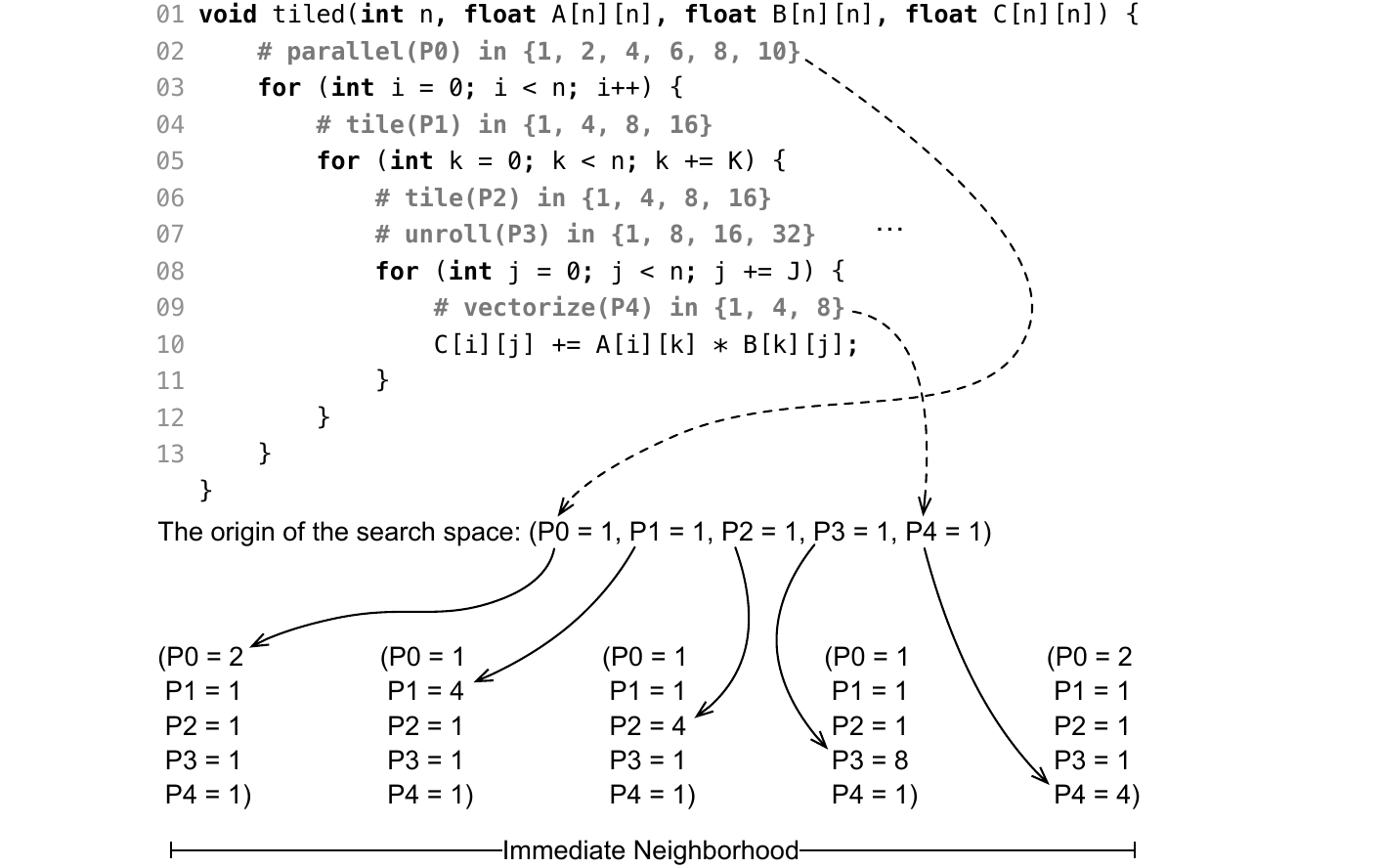}
\caption{Example of search space, with base and origin, plus an example of neighborhood.}
\Description{Example of search space, with base and origin, plus an example of neighborhood.}
\label{fig_neighborhood}
\end{figure}

\begin{example}
\label{ex_basis}
Figure~\ref{fig_neighborhood} shows the general description of all the kernels that form a five-dimensional optimization space, generated by the following optimizations:
$S = \langle \mathtt{interchange}_{i,j},\\
\mathtt{tile}_j, \mathtt{tile}_k, \mathtt{parallelize}_i, \mathtt{unroll}_j, \mathtt{vectorize}_j \rangle$.
Except for loop interchange, all these optimizations are customized via parameters.
For instance, unrolling takes in one parameter: the unrolling factor, and parallelization takes in one parameter: the number of threads.
In this example, we consider the following parameter sets:
\begin{itemize}
\item $P_0 = \mathtt{parallel}_i = \{1, 2, 4, 6, 8, 10\}$
\item $P_1 = \mathtt{tile}_k \in \{1, 4, 8, 16\}$
\item $P_2 = \mathtt{tile}_j \in \{1, 4, 8, 16\}$
\item $P_3 = \mathtt{unroll}_j \in \{1, 8, 16, 32\}$
\item $P_4 = \mathtt{vectorize}_j \in \{1, 4, 8\}$
\end{itemize}
\end{example}

A concrete kernel $k$ is defined by a choice of optimization parameters
$(p_1, p_2, \ldots, p_n)$, where each parameter belongs to a finite set of natural numbers, e.g., $p_i \in P_i \subset \mathbb{N}$.
From this observation, we define ``neighborhood'' as follows:

\begin{definition}[Neighborhood]
\label{def_neighborhood}
The {\it neighborhood} of a kernel $k = (p_1, \ldots, p_n)$ is the set of all kernels $k'$ obtained by changing exactly one parameter $p_i$ ($1 \leq i \leq n$) to an adjacent value in $P_i$, that is, $k' = (p_1, \ldots, p_i', \ldots, p_n)$ where either:

\begin{itemize}
\item $p_i' > p_i$, and for any other $p \in P_i$, if $p > p_i$, then $p \geq p_i'$;
\item $p_i' < p_i$, and for any other $p \in P_i$, if $p < p_i$, then $p \leq p_i'$.
\end{itemize}
\end{definition}

Note that the neighborhood from Definition~\ref{def_neighborhood} is a union taken over \textit{every} coordinate $i$: it contains, for each of the $n$ parameters, the kernel(s) obtained by moving that parameter to its next-larger and/or next-smaller value while holding all other parameters fixed. A neighborhood therefore mixes single-parameter perturbations drawn from all $n$ coordinates, rather than being restricted to one coordinate at a time.
Example~\ref{ex_neighborhood} illustrates this definition.

\begin{example}
\label{ex_neighborhood}
Figure~\ref{fig_neighborhood} shows the origin of the kernel optimization space, e.g.,
the vector $(p_0 = 1, p_1 = 1, p_2 = 1, p_3 = 1, p_4 = 1)$, and the neighborhood of this vector, which contains five new vectors --- one per coordinate, since each $p_i$ is already at the minimum of its domain and so only has a next-larger neighbor.
\end{example}

\subsection{Choosing the Next Kernel}
\label{sub_next_kernel}

As Definition~\ref{def_neighborhood} formalizes, a neighborhood is the set of kernels that differ from the current kernel by a change in one of its parameters, pooled across \textit{all} coordinates.
Our search moves through this neighborhood greedily: at each step, it evaluates candidates drawn from every coordinate at once and moves to whichever one improves the objective, rather than committing to a single coordinate and exhausting it before considering the others.
In this paper, we have evaluated three different techniques to choose the next kernel candidate:

\begin{enumerate}
    \item \textbf{Immediate Neighbor Evaluation}: In this approach, the algorithm evaluates the immediate neighbors (across all coordinates) and selects the best kernel that demonstrates a performance improvement over the current one. This method prioritizes quick progress, especially when the kernel neighborhood is large.
    \item \textbf{Expanded Neighborhoods}: The expanded neighborhoods approach broadens the search by considering not only immediate neighbors (first-degree) but also more distant candidates, known as higher-degree neighbors (second or third degree). This method allows the algorithm to reach beyond local configurations, helping it avoid getting stuck in suboptimal local minima by exploring a larger radius within the search space. While this flexibility can reveal better-performing configurations, it also introduces a trade-off, as examining higher-degree neighbors increases computational cost.
    \item \textbf{Shortest-Hop Exploration}: In this approach, once the hill-climbing search stops (stopping criteria are discussed in Section~\ref{sub_stop}), the search continues using the smallest value in the list of valid optimization parameters. This technique allows exploration of points that are not present in the original list of optimization parameters, as Example~\ref{ex_shortest_hop} will illustrate.
\end{enumerate}

These strategies offer a balance between search speed and thoroughness, allowing the hill-climbing algorithm to adapt to various performance surfaces, as Example~\ref{ex_next_candidate} illustrates. This example shows that our search enables the polyhedral-based optimizer to explore interactions between different optimizations that would be difficult to approximate with an exclusively static cost model.

\begin{figure}[ht]
\includegraphics[width=0.7\columnwidth]{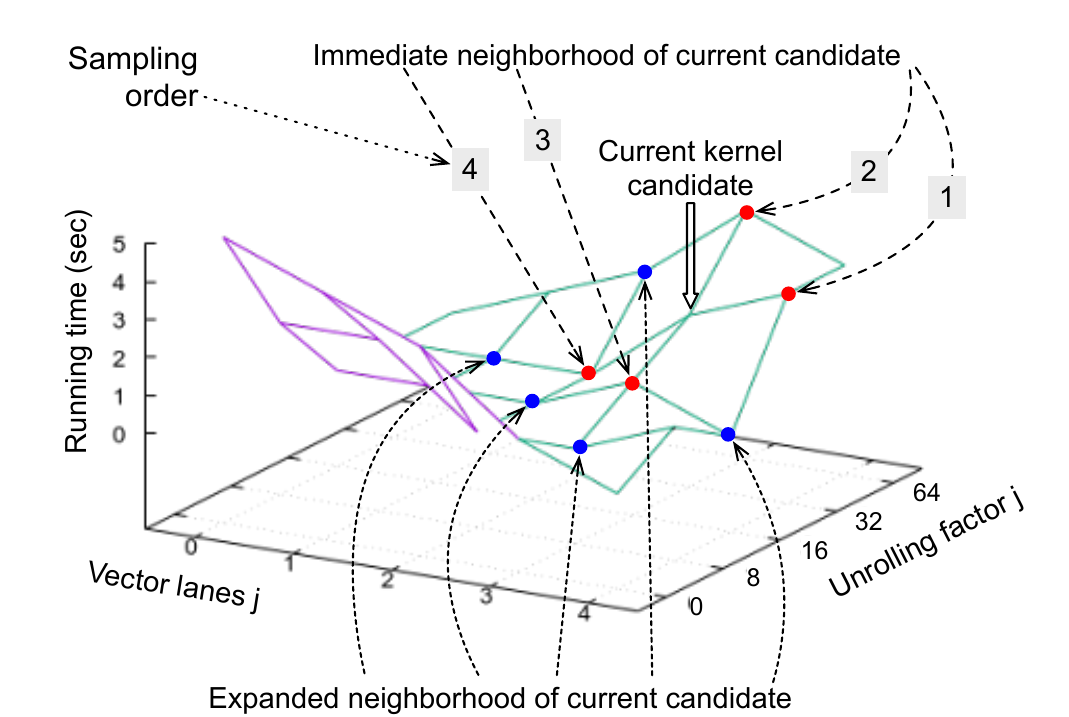}
\caption{Immediate and expanded neighborhoods of the kernel candidate on a 2D search space.
Hill climbing on the immediate neighborhood is faster, but has higher chance of becoming stuck in local minima.}
\Description{Immediate and expanded neighborhoods of the current kernel candidate.}
\label{fig_3D_neighborhood}
\end{figure}

\begin{example}
\label{ex_next_candidate}
Figure~\ref{fig_3D_neighborhood} illustrates an example of a kernel neighborhood in a 2D space shaped by two optimizations: vectorization and unrolling. The first-fit approach explores the neighboring kernels in order. In this scenario, the search would move from the current candidate to Kernel-3, even though Kernel-4 offers better performance. In contrast, the best-fit approach would select Kernel-4, as it explores the entire neighborhood. Lastly, an expanded search with a depth of 2 would examine nine kernels in total, ultimately choosing the best one.
\end{example}

The immediate and expanded neighborhoods discussed in Example~\ref{ex_next_candidate} consider a predefined list of optimization parameters and do not include new parameters formed by the additive composition of old parameters. This shortcoming is addressed in the shortest-hop variation of our search, as shown in Example~\ref{ex_shortest_hop}.

\begin{figure}[ht]
\includegraphics[width=0.7\columnwidth]{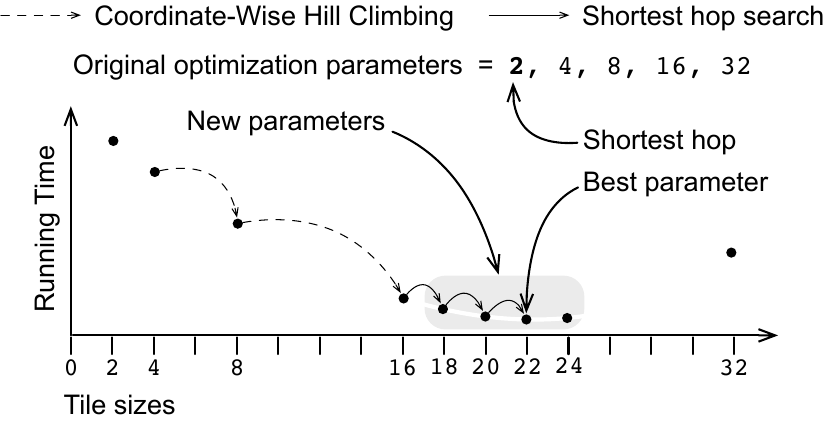}
\caption{Convergence via shortest-hop search.}
\Description{Convergence via shortest-hop search.}
\label{fig_ShortestHop}
\end{figure}

\begin{example}
\label{ex_shortest_hop}
Figure~\ref{fig_ShortestHop} illustrates how the shortest-hop search works. We assume an optimization, such as tiling, which initially admits five parameters: 2, 4, 8, 16, and 32. 
These parameters are provided by the user; however, they might not be the only valid solutions for tiling. 
Tiling windows of length 16, 18, or 20 could also be valid. 
Indeed, if any of these parameters are perfect divisors of the tiled loop, they might even be better choices, as observed by \citet{Tollenaere23}. 
In this example, let's assume that the search stops at tiles of size 16 (because trying tiles of size 32 does not yield faster kernels). 
At this stopping point, the shortest-hop search will start varying sizes by a step of two, as this is the shortest hop. 
The search still follows the same neighborhood-based procedure and evaluates tiles of length 14 and 18. 
If the latter yields faster kernels, it becomes the current candidate.
\end{example}

\subsection{Stopping Criteria}
\label{sub_stop}

The hill-climbing search stops if one of the following two conditions is met:
\begin{description}
\item[Convergence:] There is no statistical difference between the current candidate kernel $K$ and every kernel in $K$'s neighborhood.
\item[Iterations:] The search runs for a fixed number $N$ of iterations.
Each iteration consists of finding a new candidate kernel $K_i$, which is strictly faster than the previous candidate kernel $K_{i-1}$.
\end{description}
The experiments in Section~\ref{sec_eval} use a fixed bound of $N=100$ iterations.
However, in the experiments reported in Section~\ref{sec_eval}, this bound has never been achieved.
The search always stops before its $100^{th}$ iteration, due to convergence.
We determine convergence at iteration $i$ if the current candidate kernel $K_i$ cannot be shown to be faster than any kernel on its neighborhood via a Mann-Whitney U test (equivalently, the Wilcoxon rank-sum test).
We use this non-parametric test due to the small size of samples that constitute each {\it population} that it compares.
In our case, a population is the different running times that result from executing a kernel.
Our implementation runs each kernel three times, so each population being compared has three samples.

\section{Evaluation}
\label{sec_eval}

This section evaluates the following research questions:

\begin{description}
\item[RQ1:] How does our technique improve upon Pluto in terms of kernel performance?
\item[RQ2:] What is the additional time overhead introduced by our approach?
\item[RQ3:] How does our approach compare to other well-known tools in terms of kernel quality?
\item[RQ4:] How does our approach compare to other well-known tools in terms of search time?
\item[RQ5:] What is the optimal starting point for the search trajectory?
\item[RQ6:] How does the choice of the next candidate improve search efficiency and kernel quality?
\item[RQ7:] Can we generalize the adjustment approach to a different hardware model and a different optimization, such as thread blocking on a graphics processing unit?
\end{description}

\paragraph{Hardware:}
We evaluated the scheduling approaches on two architectures: x86 (Intel Xeon Max CPU 9468) and ARM-based A64FX (Fujitsu A64FX-FX700 on the Ookami Cluster). The Intel Xeon Max CPU operates at 2.6 GHz with 96 cores, and features 4.5 MiB L1d cache, 3 MiB L1i cache, 192 MiB L2 cache, and 210 MiB L3 cache. It is equipped with approximately 398 GB of RAM. The Fujitsu A64FX runs at 1.8 GHz with 48 cores and is supported by 32 GB of RAM. Each A64FX core has a 64 KB L1 cache (32 KB for data and 32 KB for instructions), and each 12-core NUMA node shares an 8 MB L2 cache. The architecture lacks an L3 cache.

\paragraph{Software:}
The ideas discussed in this paper are implemented in Pluto v0.12.0, which was the last release available at the time we performed the experiments in this section.
Henceforth, we shall use \textbf{PlutoCD} when referring to the version of Pluto augmented with our implementation of coordinate-based parameter discovery.
Pluto's standard distribution contains a collection of
kernels~\cite{Bondhugula24}.
In this section, we evaluate 11 of them.
We use every kernel that we could compile (or transform) with the following tools besides Pluto: gcc v13.2.0, clang v16.0.0, Polly release 20.0.0git, TVM v0.18.0 and IOOpt~\cite{Olivry21}.
Except for TVM, all the tools are given the same implementation of the kernel, in C++.
TVM, in turn, uses an implementation written in TensorIR, its internal domain-specific language.
The only polyhedral optimization in Pluto that lets us easily adjust parameters is tiling: in this case, tiling is parameterized by the length of the tiling window.
Therefore, each dimension of tiling gives us a dimension in the search space that we can optimize using hill climbing.

\begin{figure*}[ht]
\includegraphics[width=\linewidth]{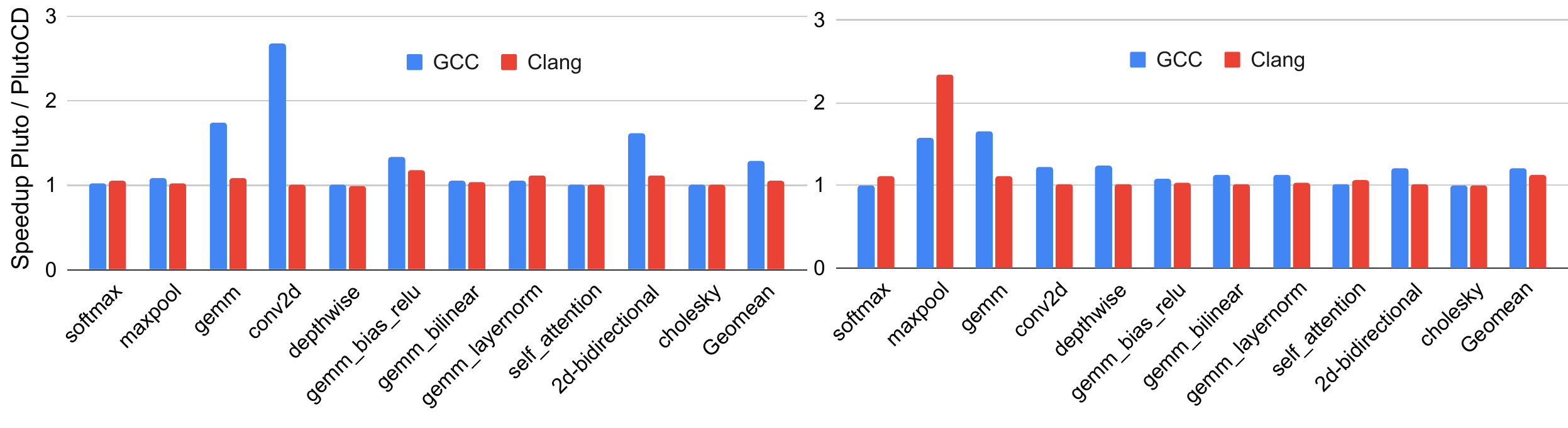}
\caption{Performance speedup of the proposed solution compared to Pluto on (a) x86 architecture and (b) A64FX architecture.}
\Description{Speed up of our solution over Pluto using (a) X86 architecture; (b) Arm A64FX architecture.}
\label{fig_kernel_performance_pluto_gcc_clang}
\end{figure*}

\subsection{Speeding Up Kernels Produced by Pluto}
\label{sub_kernel_performance_pluto}

Our first research question evaluates how much room for improvement Pluto's default choice of optimization parameters leaves.
To this end, we evaluate Pluto and PlutoCD on the eleven benchmark we have.
In this evaluation, we use two different compilers as code generators: gcc and clang, at the -O3 optimization level.
In other words, for each kernel, we optimize it using Pluto, adjust parameters using PlutoCD, and report the ratio between the running time of Pluto's kernel over PlutoCD's.

\paragraph{Discussion}
Figure~\ref{fig_kernel_performance_pluto_gcc_clang} reports our speedups. Evidently, PlutoCD always produces some speedup over Pluto. 
This observation is expected: if PlutoCD cannot optimize any parameter of the target kernel, it defaults to Pluto’s original version. 
Given our combinations of code generator (clang and gcc) and target architecture (x86 and A64FX), Figure~\ref{fig_kernel_performance_pluto_gcc_clang} reports 44 different experiments. 
In only five cases does PlutoCD settle for the same choice of parameters that Pluto uses by default (tiling windows spanning 32 loop iterations). 
Nonetheless, we still observe noticeable speedups: 1.28x on gcc/x86, 1.06x on clang/x86, 1.21x on gcc/A64FX, and 1.12x on clang/A64FX (geometric mean of ratios).

\subsection{Running Time Overhead of Hill Climbing}
\label{sub_time_overhead_pluto}

PlutoCD adds a {\it search overhead} on top of Pluto. 
This overhead includes the time required to run coordinate-based hill climbing algorithm and to sample kernels. 
Sampling a kernel requires running its implementation three times. 
The goal of this research question is to gauge the extent of this search overhead.

\paragraph{Discussion}
The table in Figure~\ref{tab_kernel_performance_pluto_searchTime} shows the search time of PlutoCD.
Without any hill climbing, we observe the following execution times for the generated code for all our eleven benchmarks (in milliseconds):
x86/gcc = 184.20; x86/clang = 2486.72; ARM/gcc = 884.89; and ARM/clang = 10829.
These are the execution time of the kernel produced by Pluto.
Once we add hill climbing to this pipeline --- comprising the time to run hill climbing and sample kernels combined with the execution time of the kernels --- we observe (time in milliseconds): 
x86/gcc = 13851; x86/clang = 107224; ARM/gcc = 64731; and ARM/clang = 428901.
Therefore, the slowdown that we obtain with PlutoCD instead of Pluto is: x86/gcc = 75.19x; x86/clang = 43.12x; ARM/gcc = 73.15x; and ARM/clang = 39.60x due to the search component added to Pluto. The improvement in the kernel's execution time is: x86/gcc = 1.23x; x86/clang = 1.06x; ARM/gcc = 1.19x; and ARM/clang = 1.04x

\begin{table}[htbp]
\centering
\caption{Time taken to generate the kernel using PlutoCD (measured in milliseconds).}
\label{tab_kernel_performance_pluto_searchTime}
\begin{tabular}{|l|r|r|r|r|}
\hline
\textbf{Processor} & \multicolumn{2}{c|}{\textbf{X86 Xeon}} & \multicolumn{2}{c|}{\textbf{ARM a64fx}} \\
\hline
\textbf{Kernel} & \textbf{GCC (ms)} & \textbf{Clang (ms)} & \textbf{GCC (ms)} & \textbf{Clang (ms)} \\
\hline
softmax           & 3806    & 5762      & 2851    & 7273    \\
\hline
maxpool            & 6900    & 117120    & 36654   & 750950  \\
\hline
gemm               & 61511   & 566470    & 298080  & 2603529 \\
\hline
conv2d             & 12159   & 177808    & 91968   & 972797  \\
\hline
depthwise          & 180429  & 2179340   & 1377427 & 7978330 \\
\hline
gemm\_bias\_relu   & 102679  & 654436    & 480187  & 5706932 \\
\hline
gemm\_bilinear     & 102908  & 548351    & 448504  & 4366545 \\
\hline
gemm\_layernorm    & 85489   & 1085021   & 457104  & 3492492 \\
\hline
self\_attention    & 35172   & 183203    & 296563  & 325822  \\
\hline
2d-bidirectional   & 11      & 60        & 38      & 179     \\
\hline
cholesky           & 2846    & 33824     & 19280   & 161574  \\
\hline
\textbf{Geomean}   & \textbf{13851} & \textbf{107224} & \textbf{64731} & \textbf{428901} \\
\hline
\end{tabular}%
\end{table}

\subsection{On the Performance of Kernels}
\label{sub_kernel_performance_tools}

There are various ways to produce executable implementations for the eleven benchmarks chosen in this evaluation, each representing different trade-offs between compilation time and executable quality. 
This section compares six approaches, each represented by a different tool: Pluto, PlutoCD, gcc -O3, Polly, IOOpt, and AutoTVM. 
The first five tools receive the same kernel implementation as input. 
In contrast, TVM receives a specification of the kernel written in its tensor expression (TVM’s DSL). 
All optimized kernels are compiled with gcc -O3 before deployment, making gcc -O3 the baseline in this study.

\begin{figure*}[ht]
\includegraphics[width=\linewidth]{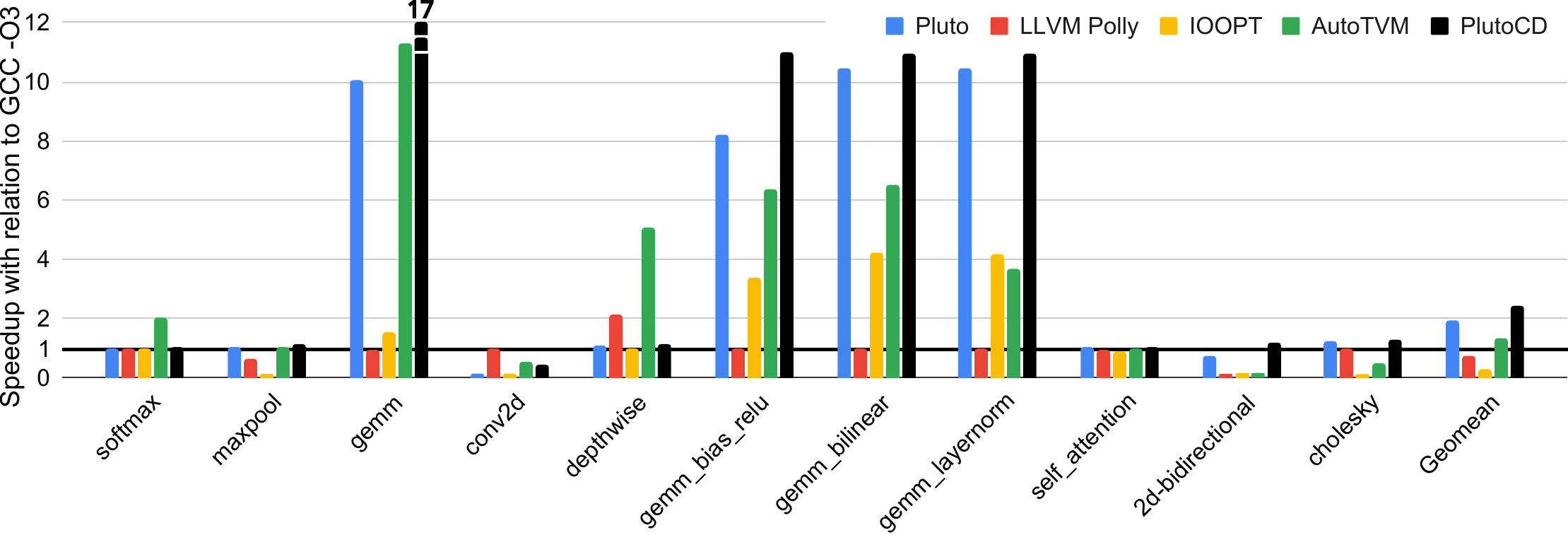}
\caption{Performance comparison between GCC -O3 and other tools on x86 architecture.}
\Description{Performance comparison between Pluto and other tools.}
\label{fig_kernel_performance_pluto}
\end{figure*}

\paragraph{Discussion}
Figure~\ref{fig_kernel_performance_pluto} compares the running time of executable kernels using gcc -O3 as the baseline.
We observe the following average running time variations (geometric mean over ratios):
x86/Pluto = 1.96x;
x86/PlutoCD = 2.46x;
x86/Polly = 0.74x;
x86/IOOpt = 0.30x;
x86/AutoTVM = 1.36x.
AutoTVM is allowed to generate and run 1,000 kernels. 
AutoTVM’s XGBoost typically converges sooner. 
TVM is only effective when templates for the input kernel are available. 
Without templates, AutoTVM may run all 1,000 samples without finding a competitive implementation. For IOOpt, we used its online interface to obtain the shape and tiling dimensions for the input kernel.
This tool uses analytical cost models, similar to the equations proposed by \citet{Narasimhan21}, to choose tile sizes.
While it allows specification of cache hierarchy parameters, the chosen tiling parameters and loop ordering did not yield competitive kernels in our x86 setting.
Our understanding is that minimizing memory traffic does not automatically optimize for vector register usage or SIMD execution. Without complementary low-level vectorization passes, IOOpt generated code might be optimal for cache movement but underutilize vector execution units.
We have reported this observation to the authors.

\subsection{Comparison of Search Times}
\label{sub_time_overhead_tools}

This paper considers two approaches to generate optimized versions of kernels: static code generation guided by a cost model (as done by Pluto, Polly, and IOOpt) and sampling (as done by AutoTVM and PlutoCD during the hill climbing phase). 
Sampling involves executing kernels and thus might impose a significant overhead on the compilation process. 
This section analyzes this overhead by comparing the search times of PlutoCD and AutoTVM. AutoTVM is allowed to run 1,000 different versions of a kernel, following the same methodology used in Section~\ref{sub_kernel_performance_tools}.

\begin{figure}[ht]
\includegraphics[width=0.65\columnwidth]{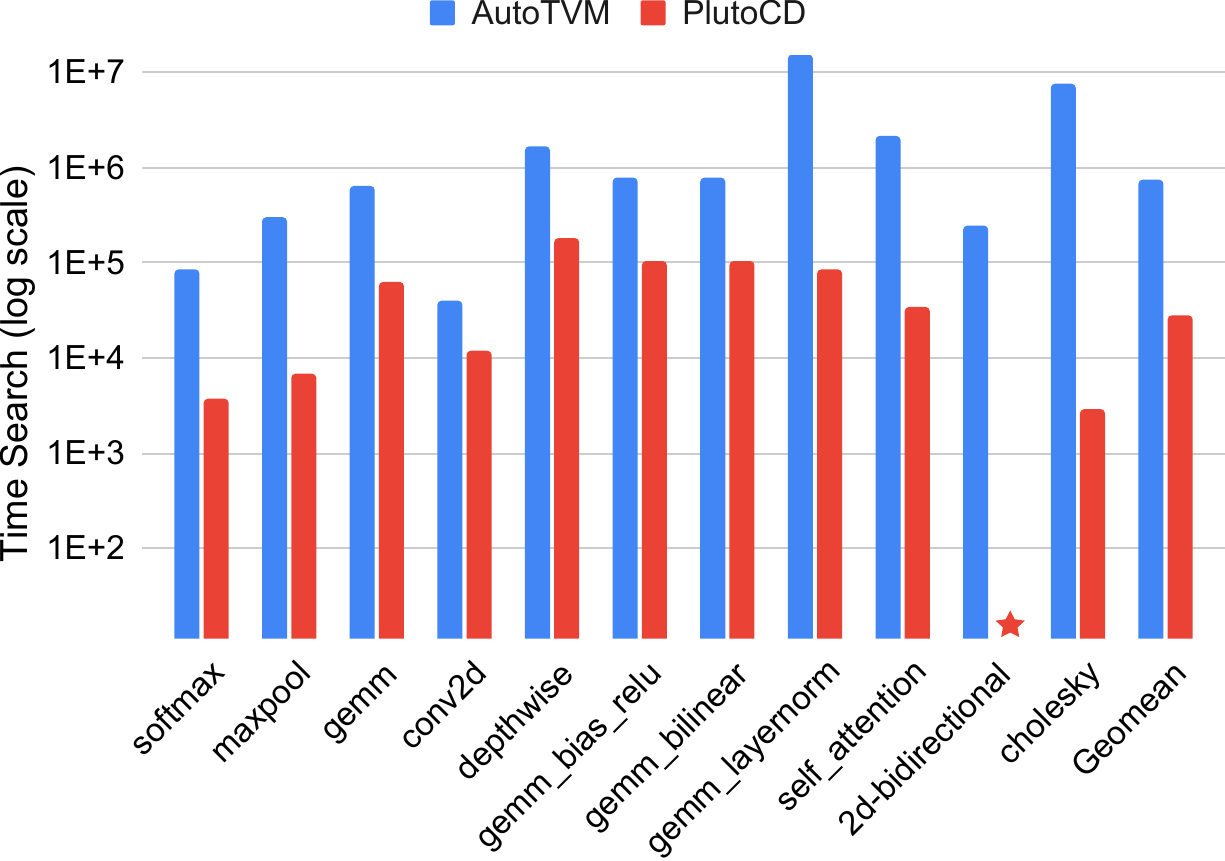}
\caption{Search time of AutoTVM and our tool (measured in milliseconds). The search time of PlutoCD on \texttt{2d-bidirectional} was too short to show.}
\Description{Search time of AutoTVM and our tool (measured in milliseconds).}
\label{fig_searchTime}
\end{figure}

\paragraph{Discussion}
Figure~\ref{fig_searchTime} compares the search times of AutoTVM and PlutoCD.
The figure shows that PlutoCD is much faster than AutoTVM: on average, based on the geometric mean of absolute running times, PlutoCD is 24.44x faster than AutoTVM in generating all eleven executables.
This difference roughly follows the difference in the number of samples collected by each framework: whereas AutoTVM compiles and runs 1,000 kernels, PlutoCD has a hard limit of 100 iterations, and it often stops before reaching half this number.
PlutoCD's advantage is further reflected in the fact that the kernels it produces are faster than those generated by AutoTVM, as discussed in Section~\ref{sub_kernel_performance_tools}.
We note, however, one limitation of this experiment: AutoTVM and PlutoCD read different inputs.
Whereas AutoTVM reads TensorIR, PlutoCD operates on C++ programs; hence, Figure~\ref{fig_searchTime} compares search times across different infrastructures.
Nevertheless, we emphasize that the difference in infrastructure alone cannot explain a two-order-of-magnitude gap between AutoTVM and PlutoCD.
To support this last statement, it is worth to compare PlutoCD with and without searches.
Post-optimization tuning causes a slowdown of 40-75x, depending on the benchmark; thus, it is the dominating factor in PlutoCD's running time.

\subsection{On the Starting Point of Coordinate-Based Hill Climbing}
\label{sub_starting_point}

Hill climbing requires a {\it seed}, or starting point. In TVM, its application is based on the ``Droplet Hypothesis''~\cite{Canesche24}, the expectation that the line between the origin in the optimization space (i.e., the unoptimized kernel) and the optimal point in this space forms a convex region. 
Following this hypothesis, hill climbing should ideally begin with a configuration where each tile window has a size of one. 
However, Pluto provides a practical starting point: a configuration where each tile window has a size of 32. 
In this section, we explore which starting point leads to the most effective implementation of PlutoCD, considering both the quality of the generated kernels and the convergence time.

\begin{table}[htbp]
\centering
\caption{Search time and kernel execution time (both in milliseconds) comparing the search starting from the origin or from Pluto's seed.}
\label{tab_seedPoint}
\begin{tabular}{|l|r|r|r|r|}
\hline
\multicolumn{1}{|c|}{\multirow{2}{*}{Kernel}} & \multicolumn{2}{c|}{From origin} & \multicolumn{2}{c|}{From seed} \\ \cline{2-5} 
\multicolumn{1}{|c|}{} & \begin{tabular}[c]{@{}c@{}}Search\\ time (ms)\end{tabular} & \begin{tabular}[c]{@{}c@{}}Kernel\\ exec (ms)\end{tabular} & \begin{tabular}[c]{@{}c@{}}Search\\ time (ms)\end{tabular} & \begin{tabular}[c]{@{}c@{}}Kernel\\ exec (ms)\end{tabular} \\ \hline
softmax & 762 & 10 & 292 & 10 \\ \hline
maxpool & 19527 & 66 & 2527 & 66 \\ \hline
gemm & 193080 & 116 & 26735 & 112 \\ \hline
conv2d & 46130 & 178 & 29969 & 211 \\ \hline
depthwise & 923478 & 3125 & 650989 & 3129 \\ \hline
gemm\_bias\_relu & 416638 & 586 & 111220 & 587 \\ \hline
gemm\_bilinear & 395848 & 586 & 79335 & 605 \\ \hline
gemm\_layernorm & 262040 & 736 & 108420 & 729 \\ \hline
self\_attention & 27140 & 425 & 19281 & 424 \\ \hline
2d-bidirectional & 98 & 0.24 & 35 & 0.24 \\ \hline
cholesky & 8698 & 260 & 14938 & 259 \\ \hline
\textbf{Geomean} & \textbf{35748} & \textbf{135} & \textbf{13893} & \textbf{137} \\ \hline
\end{tabular}%
\end{table}

\paragraph{Discussion}
Figure~\ref{tab_seedPoint} shows that beginning from Pluto’s seed (tile size 32) consistently reduces convergence time across the tested kernels compared to starting from the origin (tile size one). Specifically, starting from Pluto’s seed requires fewer iterations to reach an optimal configuration, suggesting that the search space around the seed provides a more favorable trajectory for hill climbing. When initialized from the origin, the search must explore a broader parameter range before achieving comparable performance, resulting in a slower convergence rate as the algorithm takes additional steps to approach the optimal settings found near the seed.
The starting point, however, appears to have little impact on the overall quality of the kernels produced by PlutoCD. Although Figure~\ref{tab_seedPoint} shows slight variations in the ``Kernel exec'' columns, these differences---except in the case of \texttt{conv2d}---are likely not statistically significant. In \texttt{conv2d}, beginning from the origin of the search space (in line with the Droplet Hypothesis) produces a faster kernel, which may indicate that Pluto’s seed lies outside the convex region anticipated by this hypothesis

\subsection{On the Choice of Next Candidate}
\label{sub_choice_next_candidate}

As explained in Section~\ref{sub_next_kernel}, our implementation of PlutoCD considers three different approaches to choose the next optimization parameter that hill climbing must explore. 
This research question evaluates how these different strategies for selecting the next candidate in the search trajectory affect the efficiency of convergence and the quality of the final optimized kernel produced by PlutoCD

\paragraph{Immediate Neighbor Evaluation:}
The first method discussed in Section~\ref{sub_choice_next_candidate}, the \textbf{best fit} approach, involves evaluating all the immediate neighbors of the current candidate. 
For each neighbor, we run the kernel three times and select the candidate with the best observed performance. 
Table~\ref{tab_kernel_performance_pluto_searchTime} and Figure~\ref{fig_kernel_performance_pluto} already report results for this approach. Based on those results, we know that PlutoCD’s default best fit approach, even when restricted to a neighborhood of distance one, yields kernels that are faster than those produced by Pluto alone. 
However, as the search is restricted to the immediate neighborhood, it may fail to identify configurations that require broader exploration, potentially missing global minima due to getting stuck in local minima.

\begin{figure}[ht]
\includegraphics[width=0.65\columnwidth]{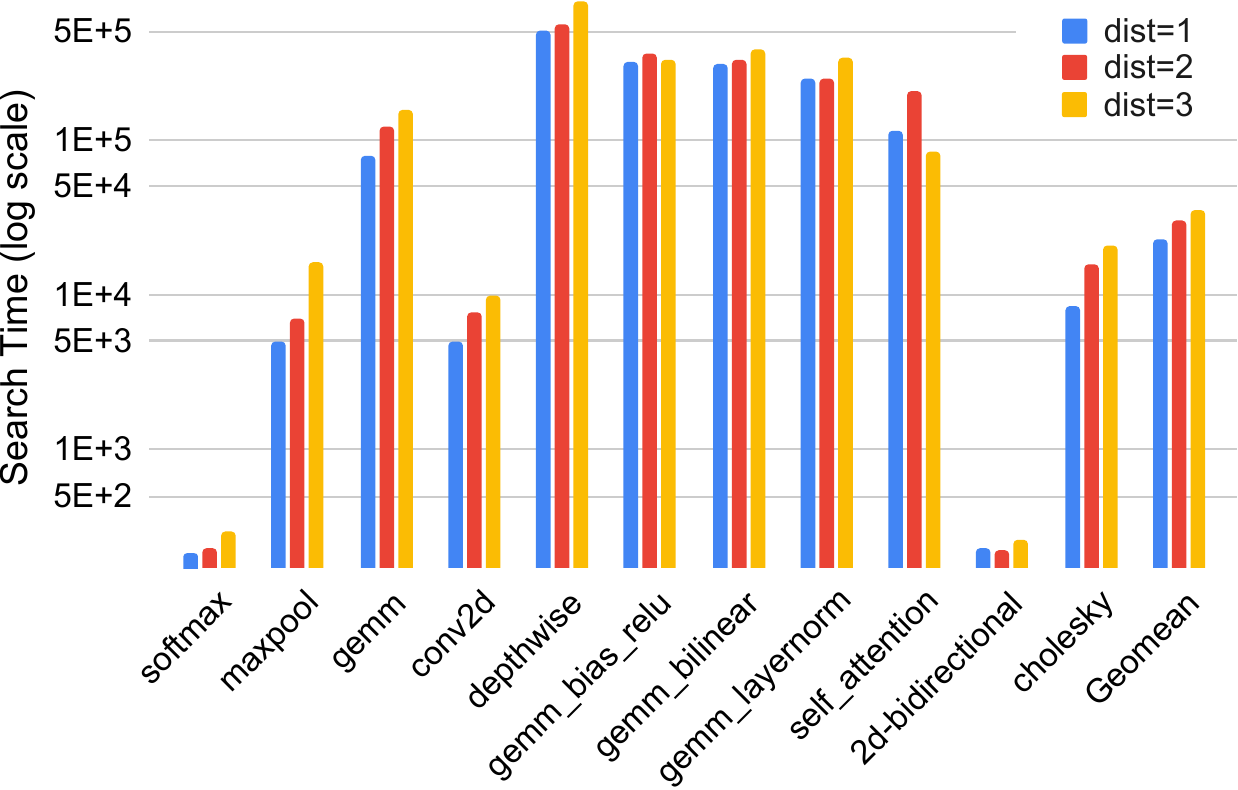}
\caption{Impact of neighborhood distance on search performance.}
\Description{Impact of neighborhood distance on the search.}
\label{fig_neighDist}
\end{figure}

\paragraph{Expanded Neighborhoods:}
To evaluate the impact of the expanded neighborhood on PlutoCD’s implementation, we evaluated neighborhoods of degree 2 and degree 3 in addition to the immediate neighborhood of degree 1 (used in Table~\ref{tab_kernel_performance_pluto_searchTime} and in Figure~\ref{fig_kernel_performance_pluto}). 
As shown in Figure~\ref{fig_neighDist}, each degree represents a wider radius of potential configurations from the starting point. These expanded neighborhoods could, in theory, allow PlutoCD to escape local minima.
However, regardless of the neighborhood degree, all three classes converged to similarly performing candidates.
This result suggests that increasing the degree of neighbors does not necessarily enhance performance. 
Instead, expanding to higher-degree neighborhoods introduces greater search-time overhead without significant improvements in kernel quality. 
Thus, limiting the neighborhood to degree 1 is typically more efficient, which is why this is the default configuration of PlutoCD.

\begin{table}[htbp]
\centering
\caption{Search time and execution time (both in milliseconds) for two distinct next-neighbor approaches.}
\label{tab_impactOpt}
\resizebox{\textwidth}{!}{%
\begin{tabular}{|l|r|r|r|r|r|r|r|}
\hline
\multicolumn{1}{|c|}{\multirow{2}{*}{Kernel}} & \multirow{2}{*}{\begin{tabular}[c]{@{}c@{}}GCC\\ -O3\end{tabular}} & \multicolumn{3}{c|}{\begin{tabular}[c]{@{}c@{}}Dynamic Hill Climbing\\ (1 degree neighbor)\end{tabular}} & \multicolumn{3}{c|}{Shortest-Hop Exploration} \\ \cline{3-8} 
\multicolumn{1}{|c|}{} &  & \begin{tabular}[c]{@{}c@{}}Search\\ Time (ms)\end{tabular} & \begin{tabular}[c]{@{}c@{}}Execution\\ Time (ms)\end{tabular} & \multicolumn{1}{c|}{Tile Sizes} & \begin{tabular}[c]{@{}c@{}}Search\\ Time (ms)\end{tabular} & \begin{tabular}[c]{@{}c@{}}Execution\\ Time (ms)\end{tabular} & \multicolumn{1}{c|}{Tile Sizes} \\ \hline
softmax & 9.9 & 8.17E+02 & 10 & (64) & 2.11E+04 & 9.8 & (488) \\ \hline
maxpool & 103 & 1.23E+04 & 57 & (4, 16) & 5.29E+04 & 57 & (16, 24) \\ \hline
gemm & 1026 & 1.82E+05 & 105 & (32,128,4) & 2.17E+05 & 91 & (256, 512, 8) \\ \hline
conv2d & 76 & 1.45E+04 & 77 & (2, 512) & 3.26E+04 & 77 & (4, 512) \\ \hline
depthwise & 3283 & 7.86E+05 & 2511 & (1,1,64,32) & 1.30E+06 & 2500 & (54,53,116,84) \\ \hline
gemm\_bias\_relu & 4022 & 1.58E+05 & 735 & (4, 64, 2) & 3.14E+05 & 720 & (10, 70, 8) \\ \hline
gemm\_bilinear & 3973 & 2.07E+05 & 593 & (16, 64, 4) & 3.04E+05 & 592 & (116,140,110) \\ \hline
gemm\_layernorm & 3977 & 2.07E+05 & 594 & (8, 64, 8) & 4.61E+05 & 587 & (264,312,252) \\ \hline
self\_attention & 1742 & 9.84E+04 & 1716 & (4, 2) & 3.80E+06 & 1619 & (502, 512) \\ \hline
2d-bidirectional & 0.25 & 1.10E+07 & 0.235 & (256) & 3.96E+08 & 0.237 & (512) \\ \hline
cholesky & 342 & 1.50E+04 & 260 & (4, 2, 2, 2) & 6.62E+04 & 259 & (2, 2, 2, 2) \\ \hline
\end{tabular}%
}
\end{table}

\paragraph{Shortest-Hop Exploration:}
The third search approach we evaluate in this section is the shortest-hop technique described in Section~\ref{sub_next_kernel} and illustrated in Example~\ref{ex_shortest_hop}.
Using this strategy, we first adjust parameters using the normal hill climbing algorithm.
Once it converges, we switch to constant variations of parameters; hence, exploiting parameter configurations that are not in the original set of predetermined configurations.
Table~\ref{tab_impactOpt} shows how this technique fares.
The shortest-hop technique yiels a marginal improvement on the running times of kernels: about 3\% on top of the default PlutoCD's immediate neighborhood.
However, its search time is substantially longer: about 30x.
Using the shortest-hop exploration, we managed to improve on the default immediate-neighborhood method in only three benchmarks: \texttt{gemm}, \texttt{gemm\_layernorm} and \texttt{self\_attention}.
Improvements naturally happen because the shortest-hop search explores more versions of the same kernel.

\subsection{On the Generalization of the Approach}
\label{sub_generalization}

The previous techniques have evaluated the impact of parameter adjustment applied onto a single compiler optimization---tiling---when producing code for a standard CPU.
This section demonstrates that this form of adjustment based on hill climbing can be applied onto a different optimization---{\it thread block tuning}---when generating code for a different architecture---the NVIDIA A100: a graphics processing unit (GPU).
In CUDA, thread block tuning refers to selecting an optimal configuration of thread blocks; specifically, the number of threads assigned to each block along the X, Y, and Z dimensions~\cite{Sethia14,Hijma23}.
This choice directly impacts memory access patterns, shared memory utilization, and overall execution efficiency.
Example~\ref{ex_thread_blocking} illustrates this optimization.

\begin{figure}[ht]
\includegraphics[width=\columnwidth]{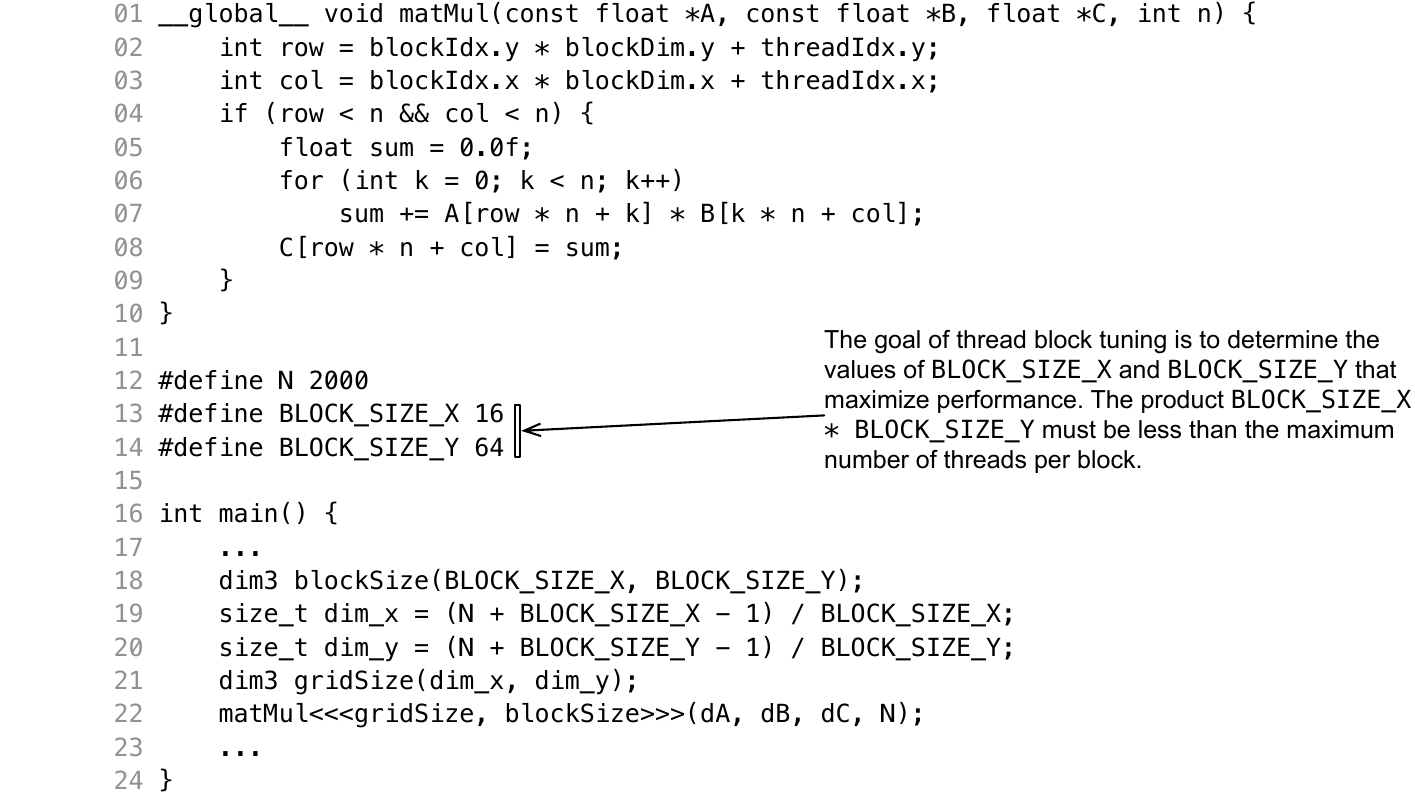}
\caption{Thread block tuning consists in determining the values of
\texttt{BLOCK\_SIZE\_X} and \texttt{BLOCK\_SIZE\_Y} that maximize performance.}
\Description{Example of thread block tuning.}
\label{fig_exampleThreadBlocking}
\end{figure}

\begin{example}
\label{ex_thread_blocking}
Figure~\ref{fig_exampleThreadBlocking} shows part of a program that implements matrix multiplication in CUDA.
Thread block tuning, in this example, involves selecting optimal values for \texttt{BLOCK\_SIZE\_X} and \texttt{BLOCK\_SIZE\_Y} to balance parallelism, memory efficiency, and hardware occupancy.
In this matrix multiplication kernel, each thread computes a single output element \texttt{C[row, col]} by iterating over a row of \texttt{A} and a column of \texttt{B}.
The number of threads per block (blockSize) is set as \texttt{(BLOCK\_SIZE\_X, BLOCK\_SIZE\_Y)}, while the number of blocks (\texttt{gridSize}) is computed to cover all $\mathtt{N} \times \mathtt{N}$ elements.
Choosing blocks that are too small leads to excessive scheduling overhead, while too large blocks may waste registers and shared memory, reducing occupancy.
The best values depend on GPU architecture, memory bandwidth, and workload characteristics.
\end{example}

\begin{figure}[ht]
\includegraphics[width=\columnwidth]{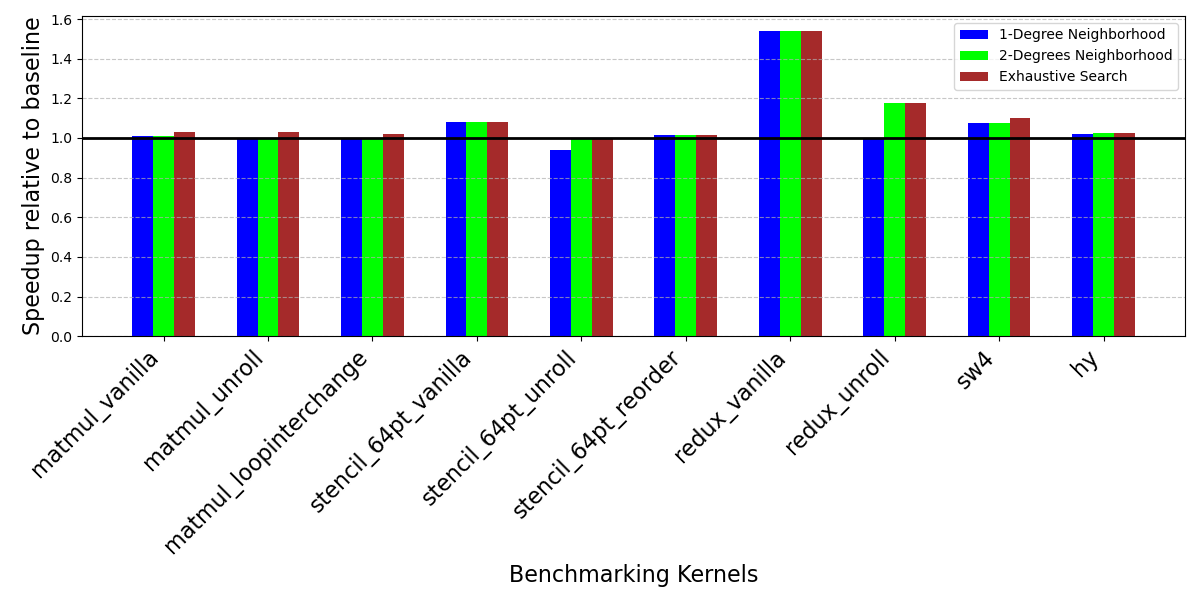}
\caption{Impact of thread block tuning on the performance of typical Cuda benchmarks. Each number is the average of three results.}
\Description{Impact of thread block tuning on the performance of typical Cuda benchmarks.}
\label{fig_gpuSpeedups}
\end{figure}

\paragraph{Benchmarks}
In this section, we evaluate the impact of post-adjustment on ten benchmarks (for the list, see Figure~\ref{fig_gpuSpeedups}).
These benchmarks were chosen to simulate the effect of compiler optimizations.
The suffix ``\texttt{\_vanilla}'' denotes the original benchmark, without any transformation.
The suffix ``\texttt{\_unroll}'' denotes the effect of loop unrolling; and the suffix ``\texttt{\_interchange}'' denotes the effect of the interchange of the two innermost loops.
Loop interchange and loop unrolling were performed manually on the original benchmark.
Two benchmarks used in Figure~\ref{fig_gpuSpeedups} were taken from an artifact made publicly available by previous work~\cite{Rawat18}.
These benchmarks simulate the application of ``statement reordering'', as proposed by \citeauthor{Rawat18}\footnote{The ``\texttt{HY}'' benchmark corresponds to the ``\texttt{hypterm}'' routine, a component of the ExpCNS Compressible Navier-Stokes mini-application developed by the U.S. Department of Energy (DoE). The SW4 and related stencil benchmarks originate from the Seismic Wave SW4 application, designed for geodynamic simulations~\cite{sw4}.}.

\paragraph{Methodology}
All the results reported in this section were observed on an NVIDIA A100 graphics processing unit, where blocks of threads have access to 1,024 physical threads.
We compare the relative performance of programs compiled with the following four thread adjustment plans:
\begin{description}
\item[Baseline:] The default thread allocation of each benchmark. This schema assigns 1,024 threads to 1D-applications (those where threads are created along one single dimension), and $32 \times 32$ threads to 2D-applications, where threads are created along the X and Y dimensions.
\item[Exhaustive:] This approach tests every possible allocation of threads along the available dimensions, varying the allocations in steps of eight threads.
\item[Degree-1:] Adjustment via hill climbing using a neighborhood of size 1.
\item[Degree-2:] Adjustment via hill climbing using a neighborhood of size 2.
\end{description}
To evaluate the impact of post-adjustment on thread block tuning, we proceed as follows: we reset to one the number of threads in all the dimensions that are subject to change.
We then apply hill climbing onto these parameters, following the methodology explained in Section~\ref{sub_next_kernel}.


\paragraph{Discussion}
Figure~\ref{fig_gpuSpeedups} shows that the adjustment phase, regardless of the technique used, is advantageous in almost every benchmark, albeit the degree of improvement varies considerably.
The geometric average running time improvement over the baseline is 5.48\%, 7.67\% and 8.53\% for hill climbing with degree one, degree two and the exhaustive search, respectively.
Hill climbing delivers results that are very close to the exhaustive search.
The extended neighborhood improves upon the neighborhood of degree one in four benchmarks: \texttt{vanilla\_matrix\_multiplication}, \texttt{sentencil\_64pt\_unroll}, \texttt{hy} and \texttt{redux\_unroll}, with a large improvement (approximately 15\%) being observed on the latter.

\begin{figure}[t!]
\includegraphics[width=\columnwidth]{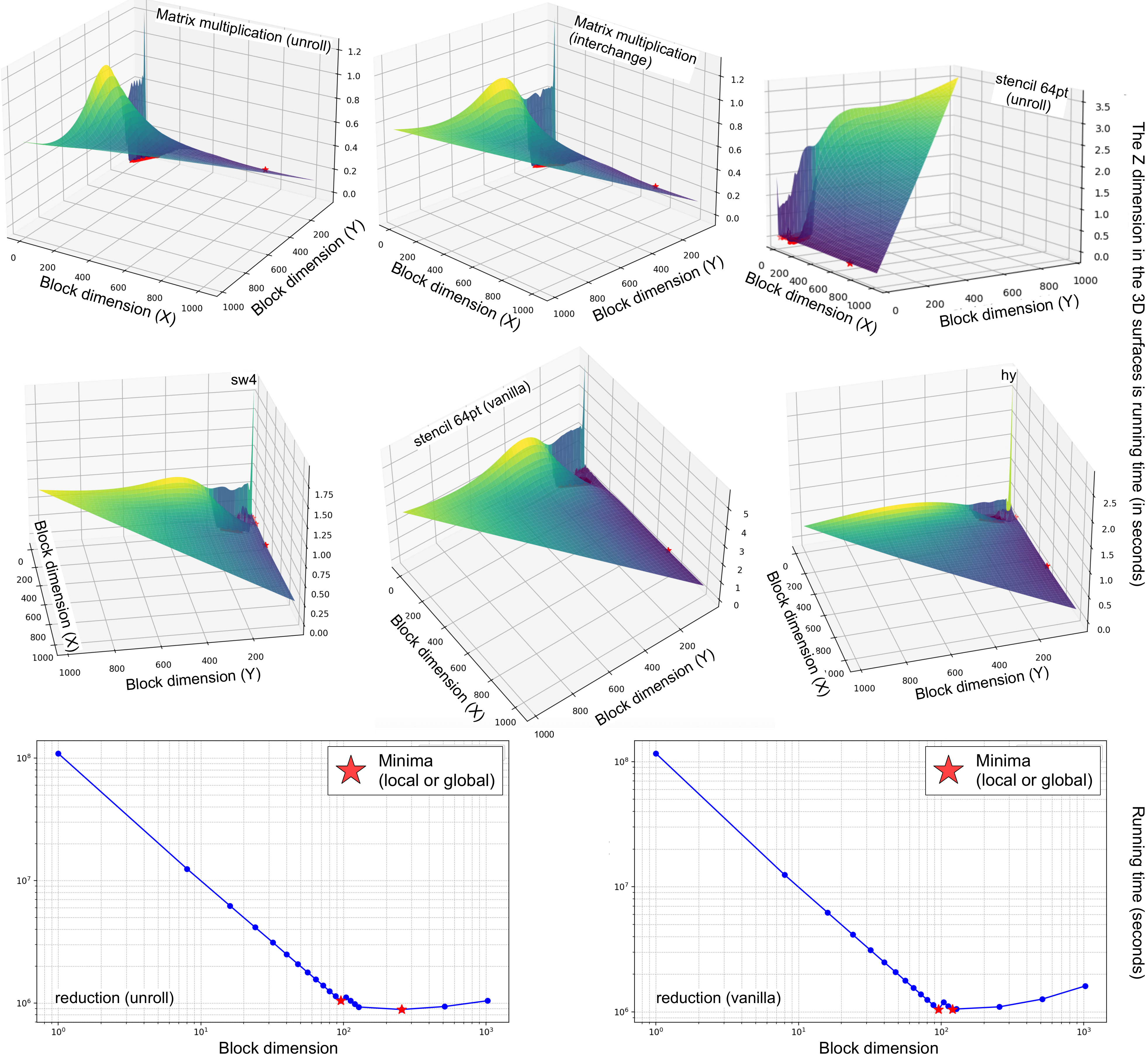}
\caption{The search space of thread blocks for seven benchmarks evaluated in this section. Each point is the average of three runs.
Red stars mark points of minima. Every surface has at least two such points.}
\Description{The search space of thread blocks for seven benchmarks evaluated in this section.}
\label{fig_blockDimensions}
\end{figure}

\begin{figure}[ht]
\includegraphics[width=\columnwidth]{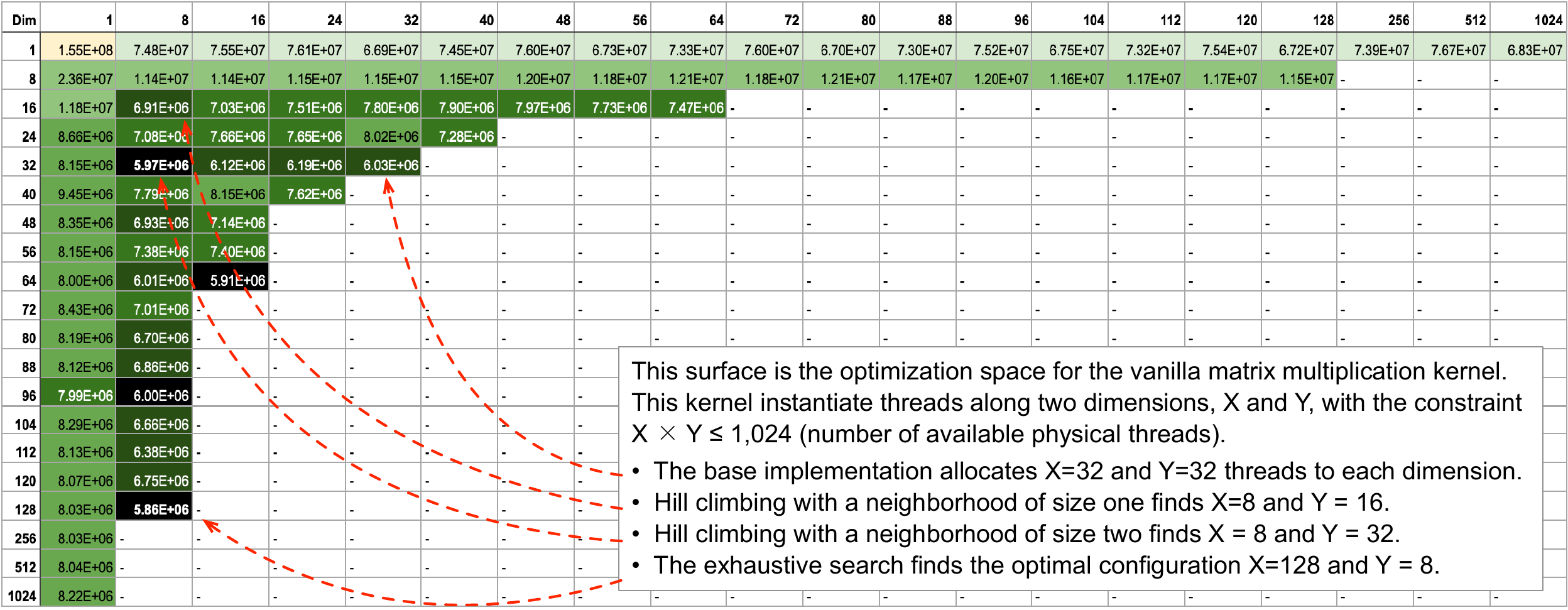}
\caption{The search space of the vanilla matrix multiplication kernel, showing the thread configurations found by different techniques.}
\Description{The search space of the vanilla matrix multiplication kernel, showing the thread configurations found by different techniques.}
\label{fig_mmSearchSpaceThreadBlocks}
\end{figure}

Convexity does not seem to be a determining factor for performance.
None of the ten benchmarks seen in Figure~\ref{fig_gpuSpeedups} yields a convex performance surface.
To demonstrate this fact, Figure~\ref{fig_blockDimensions} shows the performance surface of eight of these benchmarks.
Non-convexity surfaces might cause hill climbing stabilize at local minima.
Nevertheless, although the line of hill climbing does not traverse a convex surface, it tends to find thread configurations that are very close to the optimal performance point.
To illustrate this observation, Figure~\ref{fig_mmSearchSpaceThreadBlocks} shows the search space of the vanilla matrix multiplication kernel.
In this example, each one of the four thread-configuration approaches ends with a different configuration.
Yet, as the figure emphasizes, hill climbing, although not zeroing on the optimal configuration, is very close to it.

\section{Related Work}
\label{sec_rw}

\begin{figure}[ht]
\includegraphics[width=0.7\columnwidth]{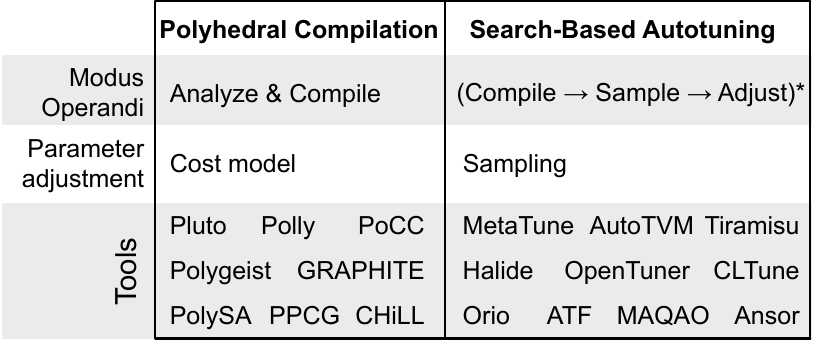}
\caption{High-level comparison between polyhedral compilation and compiler autotuners.}
\Description{High-level comparison between polyhedral compilation and compiler autotuners.}
\label{fig_Related_Work}
\end{figure}

This paper combines two code generation techniques: auto-tuning and polyhedral-based compilation.
Both share the fundamental goal of producing highly efficient code that maximizes performance by exploiting the underlying hardware's capabilities.
Additionally, both approaches focus on loop optimizations, such as tiling, unrolling, and fusion, and target performance-sensitive domains such as scientific computing, image processing, and deep learning.
The key difference between these approaches lies in how they explore the optimization space and make decisions.
Figure~\ref{fig_Related_Work} highlights these differences.

Auto-tuning, as seen in tools like MetaTune~\cite{ryu2021metatune}, AutoTVM~\cite{chen2018tvm}, Tiramisu~\cite{baghdadi2019tiramisu}, Halide~\cite{ragan2013halide}, OpenTuner~\cite{ansel2014opentuner}, CLTune~\cite{nugteren2015cltune}, Orio~\cite{hartono2009annotation}, ATF~\cite{rasch2018atf}, MAQAO~\cite{djoudi2005maqao} and Ansor~\cite{zheng2020ansor}, is a feedback-driven approach.
It generates multiple versions of code with different optimization configurations, runs these versions on the target hardware, and collects performance data, such as execution time and memory usage.
On the other hand, polyhedral-based code generation, as applied in tools like Pluto~\cite{bondhugula2008automatic, bondhugula2008practical}, Polly~\cite{grosser2012polly, pollyLLVM}, PoCC~\cite{poccSourceForge}, Polygeist~\cite{moses2021polygeist}, GRAPHITE~\cite{pop2006graphite}, PolySA~\cite{Cong18}, PPCG~\cite{verdoolaege2013polyhedral}, and CHiLL~\cite{chen2008chill}, is a theoretical model-driven approach: optimizations are guided by a cost model rather than by runtime feedback.
One primary advantage of auto-tuning is its ability to tailor optimizations to specific hardware architectures and workloads by empirically testing multiple code versions.
This adaptability allows auto-tuners to optimize code for diverse architectures, from CPUs to GPUs to FPGAs, without relying on theoretical models.
However, this trial-and-error process is computationally expensive, as running multiple code versions and gathering performance data is time-consuming.

\paragraph{Bridging the Gap}
This paper proposes a simple technique to unify polyhedral and search-based code generation.
However, previous works have already combined these techniques.
One of the earliest efforts to combine the two paradigms is \citet{Pouchet08}'s iterative polyhedral compiler, which uses the polyhedral model to prune the auto-tuner's search space.
Since that work, other research groups have explored different ways to combine the two approaches.
In this context, it is possible to place our approach relative to three distinct strategies that have been explored in the literature: (i) broader metaheuristic search over the transformation space, (ii) analytical models that eliminate runtime sampling entirely, and (iii) parametric code generation that removes the cost of recompiling between candidates.

\emph{Broader, more expensive search.} A more aggressive combination of the two paradigms is explored by \citet{Papenhausen18}'s Coding Ants framework, which extends the PPCG compiler with ant colony optimization (ACO) to jointly select scheduling algorithm choices, statement fusion/fission decisions, and tiling, modeling on the order of fifteen search-space dimensions; the underlying hypercube formulation of ACO that Coding Ants builds on is due to \citet{Blum04}. \citet{Trifunovic11} similarly studies search-based strategies for schedule and tiling selection inside a production polyhedral compiler (GRAPHITE), focusing on the scalability of search-driven compilation at compiler scale. These approaches search a substantially larger and more general space than ours — the full scheduling and fusion decision space, not just the numeric parameters of an already-chosen kernel shape — and can consequently find configurations our technique cannot reach. This generality comes at a cost: population-based metaheuristics such as ACO require many iterations over many candidate colonies, each colony member requiring its own kernel execution, which is considerably more expensive than the handful of samples our hill-climbing search needs per kernel (Section~\ref{sec_eval}). Our contribution is not a more thorough search; it is showing that, once a polyhedral compiler has already committed to a kernel shape, a lightweight, model-free local search over that shape's numeric parameters recovers most of the available performance at a small fraction of this cost.

\emph{Analytical, zero-sampling models.} A separate line of work avoids runtime sampling altogether by deriving tile sizes (and, in some cases, fusion decisions) from an analytical cost model. \citet{Sarkar00} give an early cache-miss-based analytical model for tile size selection. \citet{Narasimhan21} propose a fast, general tile size selection model that computes tile sizes as the zeros of a low-degree polynomial capturing temporal and spatial reuse, incurring essentially no compile-time overhead, and demonstrate it across linear algebra, DSP, and image-processing kernels. \citet{Jangda18} jointly model fusion and tile size selection analytically for image-processing pipelines expressed in PolyMage. \citet{Jayaweera24} extend this line of analytical modeling to a different objective: energy rather than running time for GPU kernels.
More recently, \citet{Tollenaere23} used similar techniques to speed up Ansor's empirical search, avoiding evaluation of kernel versions that are unlikely to yield good performance.
These techniques have a real advantage our approach does not share: zero runtime sampling cost, in contrast to the search overhead we report in Section~\ref{sub_time_overhead_tools}. Their limitation is the difficulty to model hardware characteristics. Modern hardware is very complex, and often lacks open specifications that model developers can use. As we show in this paper, hill-climbing adaptation can approximate hardware characteristics without analytical models, at the price of empirical sampling. The two ideas are complementary rather than competing: an analytical tile size model such as \citet{Narasimhan21}'s could replace Pluto's fixed default as the seed for our search, likely reducing the number of hill-climbing iterations needed to converge.
We leave this composition to future work.

\emph{Parametric, recompilation-free code generation.} A third line of work targets a different cost than either of the above: the overhead of regenerating and recompiling code for every candidate parameter value during search. \citet{Hartono09} generate parametric multi-level tiled code (including for imperfectly nested loops) in which tile sizes are runtime rather than compile-time constants, so different candidates can be evaluated without recompilation. DynTile~\cite{Hartono10} pursues the same goal for parallel execution on multicore processors. This directly addresses a limitation of our approach that we quantify in Section~\ref{sub_time_overhead_tools}: each hill-climbing step in our implementation requires regenerating and recompiling the kernel, which is the dominant source of the reported $40$--$75\times$ search-time overhead relative to Pluto alone. Adopting a parametric code generation strategy is a natural way to reduce this overhead in future versions of our approach without changing the hill-climbing search itself.

We believe that using hill climbing to adjust the parameters of polyhedral-based optimizations, seeded exclusively by a static, model-free polyhedral analysis, is an original contribution of this paper.
Hill climbing in autotuning itself is not new: the droplet search algorithm was designed to speed up AutoTVM's search time~\cite{Canesche24}.
Variants of this algorithm were subsequently applied to Ansor~\cite{Li24} and TVM's MetaScheduler~\cite{Canesche24b}.
The key difference in this work is that previous uses of hill climbing were in tandem with autotuners, with initial spaces chosen via human intervention~\cite{Canesche24} or other autotuning algorithms like evolutionary search~\cite{Canesche24b}.
This paper exclusively uses static analyses based on the polytope model to determine the search space for hill climbing.

\section{Conclusion}
\label{sec_con}

This paper has introduced a methodology to adjust the optimization parameters used by polyhedral-based compilers, making them more competitive with the autotuners that have gained popularity in recent years.
The approach proposed in this paper uses hill climbing to fine tune the parameters of static compiler optimization. The strength of this method lies in observing that a simple hill-climbing tuning approach, once combined with static optimizers, such as Pluto, can bring these optimizers closer to more extensive autotuners, such as AutoTVM, for instance.
This combination of polyhedral optimization with hill climbing-based parameter adjustment is synergistic in a number of ways.
On one hand, the polyhedral-based compiler provides hill climbing with a feasible optimization space to explore. On the other, the feedback-driven adjustment phase makes the compiler more attuned to the finer details of the target hardware. Without the optimization space generated by the polyhedral compiler, hill climbing would lack direction as it would have no clear space to navigate. Conversely, without the refinement phase of hill climbing, effectively using the polyhedral-based compiler would be challenging, as it would not have the necessary adjustments to align its static optimization parameters with the complexities of the target architecture---a task too intricate to be captured by a general cost model.

\paragraph{Software}
An artifact containing all the tools necessary to reproduce the experiments in Section~\ref{sec_eval} is available at \url{https://github.com/xintin/kelpie}.

\bibliographystyle{ACM-Reference-Format}
\bibliography{references}

\end{document}